\documentclass[preprint,12pt,3p]{elsarticle}

\usepackage{amsmath}
\usepackage{amssymb}
\usepackage{amsfonts}
\usepackage{listings,xcolor,caption, mathtools, wrapfig}
\usepackage{subfig}
\usepackage{float}
\usepackage{graphicx}
\usepackage{multicol}
\usepackage{tcolorbox}
\usepackage{multirow}
\usepackage{enumerate}
\usepackage[normalem]{ulem} 
\usepackage{siunitx} 
\usepackage{booktabs}
\usepackage[mathscr]{euscript}
\usepackage{makecell}

\renewcommand{\vec}{\textbf}

\journal{Computer Methods in Applied Mechanics and Engineering}

\begin{document}

\begin{frontmatter}

\title{Physics-informed neural networks for blood flow inverse problems}


\author[label1,label2,label4]{Jerem\'ias Garay}
\author[label4,label5,label6,label9]{Jocelyn Dunstan}
\author[label2,label3,label4,label9]{Sergio Uribe}
\author[label1,label4,label8]{Francisco Sahli Costabal}

\cortext[cor1]{Corresponding author}
\ead{fsc@ing.puc.cl}

\address[label1]{Department of Mechanical and Metallurgical Engineering, Pontifical Catholic University of Chile}
\address[label2]{Center of Biomedical Imaging, Pontifical Catholic University of Chile}
\address[label3]{Department of Medical Imaging and Radiation Sciences, Monash University}
\address[label4]{Millennium Institute for Intelligent Healthcare Engineering (iHealth), Chile}
\address[label5]{Department of Computer Science, Pontifical Catholic University of Chile}
\address[label9]{Center for Mathematical Modeling, University of Chile}
\address[label6]{Institute for Mathematical Computing Engineering, Pontifical Catholic University of Chile}
\address[label8]{Institute for Biological and Medical Engineering, Pontifical Catholic University of Chile}
\address[label9]{Department of Radiology, Pontifical Catholic University of Chile}

\begin{abstract}
Physics-informed neural networks (PINNs) have emerged as a powerful tool for solving inverse problems, especially in cases where no complete information about the system is known and scatter measurements are available. This is especially useful in hemodynamics since the boundary information is often difficult to model, and high-quality blood flow measurements are generally hard to obtain. In this work, we use the PINNs methodology for estimating reduced-order model parameters and the full velocity field from scatter 2D noisy measurements in the ascending aorta. The results show stable and accurate parameter estimations when using the method with simulated data, while the velocity reconstruction shows dependence on the measurement quality and the flow pattern complexity. The method allows for solving clinical-relevant inverse problems in hemodynamics and complex coupled physical systems.  
\end{abstract}


\end{frontmatter}


\section{Introduction}
\label{sec:introduction}

Computational hemodynamics has been established as a relatively new research area in which blood flow is studied in many scenarios. Applications include the blood ejected from the heart through the aorta artery or blood flow through small vessels, such as the capillaries in the brain, or the flow through the coronaries arteries. Moreover, hemodynamics simulations have proven helpful for planning and assessing different cardiovascular pathologies, for instance, aortic dissection \cite{erbel2001diagnosis}, stenosis \cite{carabello2009aortic}, and aneurysms \cite{sakalihasan2005abdominal}, which are known to be highly affected by blood flow patterns and can generally be life-threatening and progressive. In a clinical context, the main reason for engaging in simulations in a complex cardiovascular scenario is to gain prediction capabilities, a feature that cannot be attained only based on images.

However, one of the main limitations of hemodynamical simulations and their clinical use is the strong sensitivity of the results to the modeling assumptions and parameters. Different imaging techniques have helped to fill this gap, helping to obtain patient-specific geometries and boundary condition information, which is essential for most physical models. Magnetic Resonance Imaging (MRI) \cite{mcrobbie2003mri} and Computational Tomography (CT) \cite{fleischmann2016computed} are widely used and can be obtained relatively easily for small to large patient cohorts and healthy volunteers. Moreover, tissue and blood flow motion is usually measured by the so-called Phase-Contrast MRI (PC-MRI) technique \cite{hom2008velocity}, capable of encoding the tissue velocity into the phase of the emitted signal. From this method, we can obtain time-resolved vascular measurements, usually consisting of 20 to 30 snapshots within the cardiac cycle. Although the temporal and spatial resolution on different applications have become progressively better, some physical properties cannot be measured using purely non-invasive techniques, such as the stress at the boundary of the vessels, namely the wall shear stress, pressure drops across a stenotic region of the vessel, and other mechanical properties of the arteries.

Another limitation of hemodynamic simulations is the large computational cost and time required to make the models realistic. For this reason, several strategies for reducing model complexity have been proposed. For instance, the so-called \emph{reduced order} modeling approaches, which try to simplify some physical phenomena, making the simulation realistic while keeping a tractable computational time and needed resources \cite{nolte2022inverse, caiazzo2018mathematical}. In hemodynamics, the range of the applications could vary in size, making the reduced strategy also change. Some examples on different applications are in: the coronary arteries \cite{kim2010patient, grande20221d, bertoglio2019junction}, the aorta artery \cite{bertoglio2013fractional, ismail2013adjoint}, the pulmonary artery \cite{marcinno2023computational} and heart-valve problems \cite{piersanti20223d, pase2023parametric}. 

A popular choice for medium-to-large size arteries is using the three-element Windkessel model \cite{windkesselartery}. This model considers the vessels' compressibility and ability to store elastic energy during systole to release it in diastole. In general terms, a Windkessel model introduces a 0D differential equation at the vessel outlet, modeling the pressure and flow with an equivalent electric circuit, where the blood flow is represented by the current and the pressure as the voltage \cite{franck1990basic}. The model also introduces a resistance parameter, where higher resistance values can explain difficulties in the blood flow through that specific portion of the artery. Additionally, the model introduces a compliance $C$, which considers the vessel's ability to store elastic energy.

The clinical value of estimating the Windkessel parameters on a specific data set relies on the physical information one can obtain from them, such as localized flow resistance effects or to know if a specific vessel presents a reduced elastic response. This data-driven information could be seen as a new cardiovascular biomarker for later clinical patient assessment due to its specificity and non-invasive nature. In this line, different strategies have been used to infer Windkessel parameters from partial measurements to make blood flow simulations patient-specific. For example, Arthurs et al. \cite{arthurs2020flexible} applied the Reduced-Order Unscented Kalman Filter to the 3D time-dependent Navier–Stokes model with simplified fluid-solid interaction effects on the wall to estimate the Windkessel parameters in the boundary conditions. Using variational data assimilation, Fevola et al. \cite{fevola2021optimal} estimated a single resistant model (i.e., $C=0$) on a stationary Stokes problem. Bertoglio et al. \cite{bertoglio2012sequential} used a Kalman Filter approach to estimate elastic and Windkessel parameters of the vessels from wall displacement measurements. Finally, Garay et al. \cite{garay2022parameter} used a novel data assimilation term in their sequential inverse problem workflow to estimate three-element Windkessel parameters from highly aliased and noisy PC-MRI blood flow measurements.


However, all the works mentioned above strongly depend on the complete description of the physical model, sometimes leading to oversimplified solutions, especially in cases where the physics is complex and poorly understood. For that reason, in this work, we propose to use physics-informed neural networks (PINNs) \cite{raissi2019physics, karniadakis2021physics}, which have shown great potential in solving inverse problems where the physical information is not complete, having the networks the ability to learn or to discover the missing parts from the data itself. We will use a feed-forward network architecture to represent the flow state of the coupled Navier-Stokes equations with the Windkessel model. Note that partial knowledge of the boundary conditions is very common in hemodynamics and usually leads to ill-posed problems which classical approaches failed to address \cite{nolte2022inverse}. Hence, PINNs allow us to estimate the time-dependent 3D velocity and pressure field from scatter-simulated medical images by solving the forward problem and the closest set of Windkessel parameters by solving the inverse problem simultaneously.

PINNs have emerged as a novel methodology for solving complex forward problems and ill-posed inverse problems in many research areas, such as in fluid mechanics \cite{cai2021physics, xiao2020flows,mathews2021uncovering}, electrophysiology problems \cite{sahli2020physics,ruiz2022physics}, geosciences and wave-propagation problems \cite{li2021physics, maharjan2022deep}, non-linear solid mechanics \cite{tac2023benchmarks, BAZMARA2023152, haghighat2020deep} and also in cardiovascular biomechanics \cite{arzani2021uncovering, kissas2020machine, arzani2022machine, SUN2020112732}. See Ref \cite{cuomo2022scientific} for an extensive review of all current applications. Although PINNs have been applied in many problems, and the Navier-Stokes equations have been focus of PINNs since its inception \cite{raissi2019physics}, to the best of the author's knowledge, this is the first time that PINNs have been applied in a coupled 3D fluid mechanics problem to estimate hemodynamic parameters from medical images. 

The rest of this article is organized as follows. In Section \ref{sec:methods}, we present the mathematical model, the neural architecture to be used, and the reference velocity and simulated measurements used for the inverse problem. In Section \ref{sec:results}, we present the results obtained for both studied regimes: the steady and transient flow cases. A discussion of the results and future lines of work are given in Section \ref{sec:discussion}. Finally, the conclusions are presented in Section \ref{sec:conclusions}.

\section{Methods}
\label{sec:methods}
We start by introducing the model that describes the physics of blood flow in arteries.
\subsection{The mathematical model}
\label{sec:mathematical_model}
\begin{figure}[!hbtp]
\centering
  \includegraphics[trim=70 0 250 0, clip, width=0.33\textwidth]{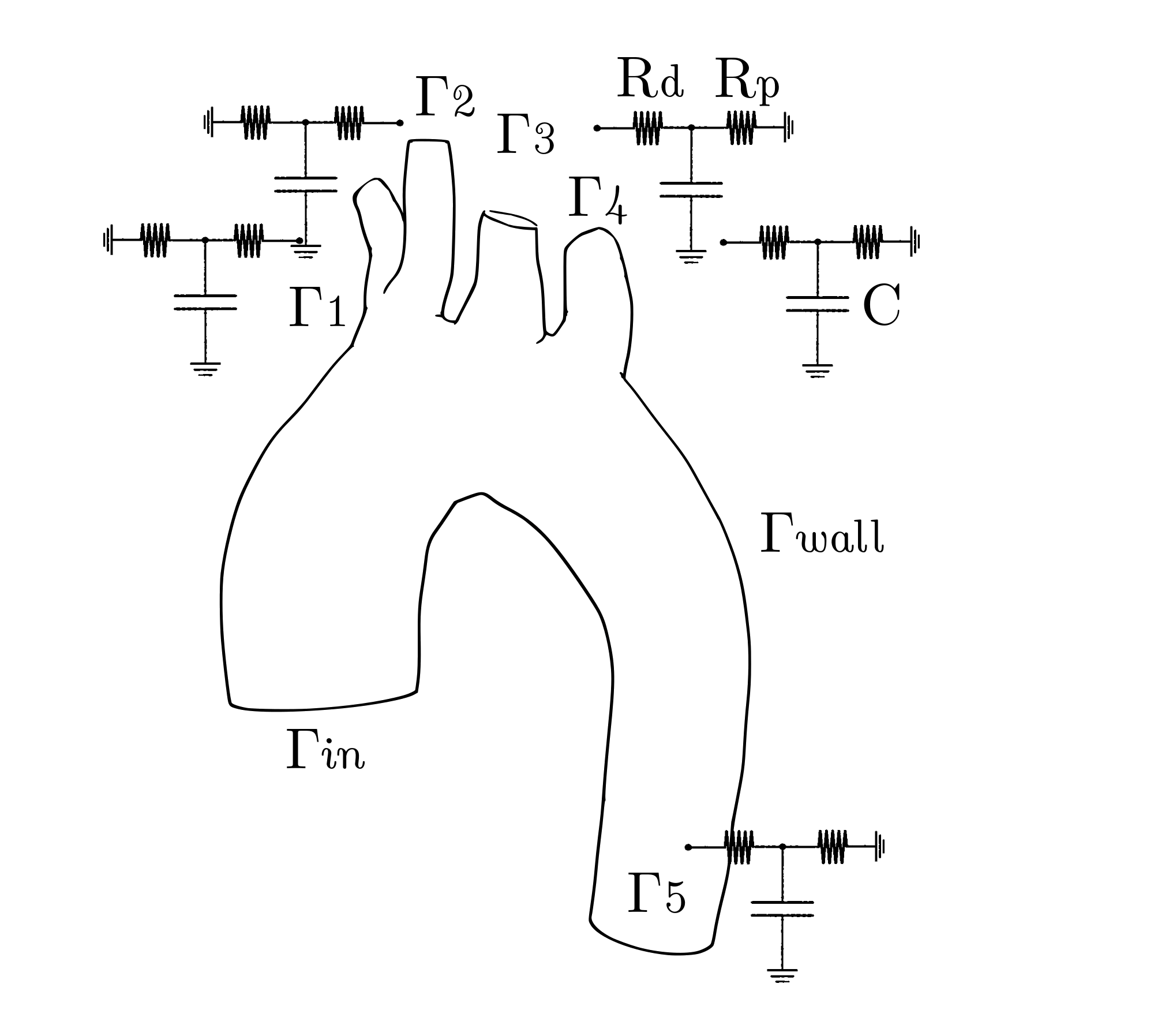} 
  \caption{Aortic domain $\Omega$. The Windkessel outlets are defined at $\Gamma_1$ to $\Gamma_5$ consisting of 2 resistors ($R_d$, $R_p$) and 1 capacitor $C$ each.}
\label{fig:meshes}
\end{figure}

Let $\Omega \subset \mathbb{R}^3$ be the interior (lumen) of the thoracic aorta, represented in Figure \ref{fig:meshes}, with its boundary $\partial \Omega$ sub-divided as follows: 
$$
\displaystyle
\partial \Omega = \Gamma_{in} \cup \Gamma_{wall} \cup \big( \displaystyle \cup_{k=1} \Gamma_k \big) , $$
 where $\Gamma_{in}$ is the inlet boundary, $\Gamma_{wall}$ the arterial wall, and $\Gamma_1,\dots,\Gamma_K$ are the $K$-outlet boundaries. We consider that the blood flow in this domain is governed by the incompressible Navier-Stokes equations, with velocity $\vec{u}(\vec x,t)$ and pressure $p(\vec x,t)$:
\begin{equation}
\label{eq:NS}
\begin{cases}
\displaystyle \rho \frac{\partial \vec{u}}{\partial t} + \rho \big ( \vec{u}  \cdot \nabla \big) \vec{u} - \mu \Delta \vec{u} + \nabla p  =  0 \quad \text{in} \quad \Omega \times [0,T], \\[0.2cm]
\nabla \cdot \vec{u}  =  0 \quad \text{in} \quad \Omega \times [0,T], \\[0.2cm]
\vec{u}  =  \vec{u}_{inlet} (t) \quad \text{on} \quad \Gamma_{in} \times [0,T], \\[0.2cm]
\vec{u} = \vec 0 \quad \text{on} \quad \Gamma_{wall} \times [0,T], \\[0.2cm]
\displaystyle \mu \frac{\partial \vec{u}}{\partial n} - p\vec{n} = -P_k(t)\displaystyle\vec{n} \quad \text{on} \quad \Gamma_k \times [0,T], \quad k = 1,\dots,K
\end{cases}
\end{equation}
 where $\rho$ is the blood density and $\mu$ the dynamic viscosity of the fluid. Note that a Dirichlet inflow and a non-slip boundary conditions are adopted at $\Gamma_{in}$ and $\Gamma_{wall}$, respectively. Moreover, the pressure at the outlet $P_k(t)$ is assumed given by the \textit{three-element Windkessel} model:
\begin{equation}
\label{eq:WK}
\begin{cases}
P_k = R_{p,k} \ Q_k + \pi_k , \\[0.2cm]
Q_k  = \displaystyle \int_{\Gamma_k} \vec{u} \cdot \vec{n} , \\[0.2cm]
\displaystyle C_{k} \frac{d \pi_k}{dt} + \frac{\pi_k}{R_{d,k}} = Q_k .
\end{cases}
\end{equation}

In the Windkessel model, $R_{p,k}$ and $R_{d,k}$ represent the resistance of the vasculature proximal and distal to $\Gamma_k$, respectively, while $C_{k}$ is the compliance of the distal vessels. The exterior normal vector of the outlet is represented by $\vec{n}$. 

In the special case of a stationary flow, all time derivatives vanish from both models, resulting in the lower dimensional system an effective \emph{one-element} Windkessel boundary condition with no internal pressure variable $\pi$. The equations of the Windkessel model are reduced to:
\begin{equation}
\label{eq:WK_sta}
\begin{cases}
P_k = (R_{p,k} + R_{d,k}) \ Q_k = R_{t,k} \ Q_k, \\[0.2cm]
Q_k  = \displaystyle \int_{\Gamma_k} \vec{u} \cdot \vec{n}. \\
\end{cases}
\end{equation}
with the distal and proximal resistances absorbed in a total resistance parameter $R_t$. 

\subsection{Physics-Informed Neural Networks (PINNs)}

\begin{figure}[h]
\centering
  \includegraphics[trim=0 0 0 0, clip, width=1.0\textwidth]{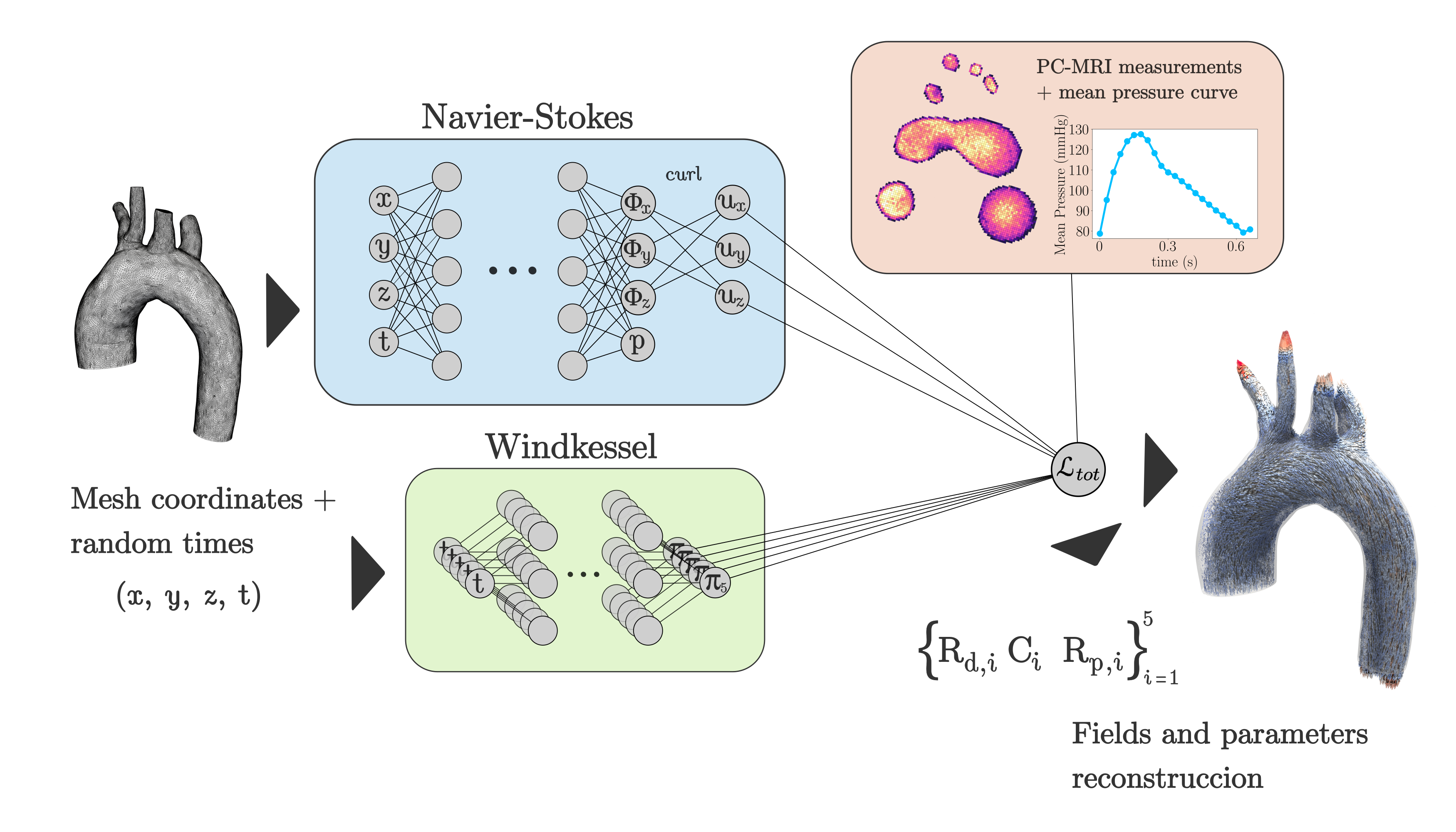}
  \caption{Feed-forward neural network architecture used for the transient problem. The network represents a vector potential $\Phi$ and the pressure field. After physics-informed training, the output of the network is used for reconstruct the 3D velocity field and the Windkessel parameters of all the domain outlets. }
\label{fig:NN}
\end{figure}

We will use PINNs to solve the inverse problem of inferring the Windkessel model parameters $R_{p,k},R_{d,k},C_k \quad \forall k=1,...,K$, and the velocity and pressure fields in the aorta from partial velocity measurements $\vec{u}_{meas}$, obtained from PC-MRI and an average pressure measurement in the outlets $\Bar{p}_{meas}$. We represent the solution of the coupled system in Equation~\ref{eq:NS} and \ref{eq:WK} by a neural network consisting of $n$ feed-forward layers, as shown in Figure \ref{fig:NN}:
\begin{equation}
    \big ( \vec{u} (\vec{x},t), \ p(\vec{x},t) \big )= \mathscr{NN} (\vec{x},t;\mathbb{W}, \vec{b})
\end{equation}
Where the neural network $\mathscr{NN}$ outputs the fields $\vec{u}$ and $p$, and takes as input the coordinate points in space and time, parametrized by weights ($\mathbb{W}$) and biases ($\vec{b}$). The values for $\mathbb{W}$ and $\vec{b}$ are obtained after training the network. To achieve this, we define a loss function $\mathscr{L}_{tot}$ that encourages learning the measurements, the normalized physics equations \eqref{eq:NS}a, \eqref{eq:NS}b and, \eqref{eq:WK}, and the non-slip boundary condition in 
\eqref{eq:NS}d:
\begin{equation}
    \mathscr{L}_{tot} =  \mathscr{L}_{NS} + \mathscr{L}_{WK} + \mathscr{L}_{BC} +  \mathscr{L}_{u,data} +  \mathscr{L}_{p,data} + \mathscr{L}_{gradp}.
\end{equation}
where the individual loss components are defined as:
\begin{eqnarray}
 \label{ec:loss_terms}
 \label{ec:loss_ns}
 \mathscr{L}_{NS} & = &  \lambda_{NS} \ \bigg |\bigg | \displaystyle \frac{\partial \vec{u}}{\partial t} +  \big ( \vec{u}  \cdot \nabla \big) \vec{u} - \frac{1}{Re} \Delta \vec{u} + \nabla p     \bigg |\bigg |_{\Omega}   +  \lambda_{NS} \ \big |\big | \nabla \cdot \vec{u}  \big |\big |_{\Omega}, \\
 \label{ec:loss_wk}
  \mathscr{L}_{WK} & = & \lambda_{WK} \ \sum_{k=1}^{K} \big | \big | p_{k} - R_{p,k} Q_{k} - \pi_{k}  \big | \big |_{\Gamma_{k}} +  \big | \big |   C_{k} \frac{d \pi_k}{dt} + \frac{\pi_k}{R_{d,k}} - Q_k  \bigg | \bigg |_{\Gamma_{k}}, \\[0.2cm]
 \mathscr{L}_{BC} & = & \lambda_{BC}  \ || \ \displaystyle  \vec{u}    \ ||_{\Gamma_{wall}}, \\
 \label{ec:loss_data}
 \mathscr{L}_{u,data} & = & \lambda_{data} \ || \ \displaystyle  \vec{u} - \vec{u}_{meas}   \ ||_{\Omega_{meas}}, \\
  \label{ec:loss_udata}
   \label{ec:loss_data_p}
 \mathscr{L}_{p,data} & = & \lambda_{data} \ || \ \displaystyle  \Bar{p} - \Bar{p}_{meas}   \ ||_{\Omega_{meas}}, \\
  \label{ec:loss_gradp}
 \mathscr{L}_{p,gradp} & = & \lambda_{gradp} \sum_{k=1}^K \big |\big | \ \nabla p  \ \big |\big |_{\Gamma_k}.
\end{eqnarray}

The last added term \eqref{ec:loss_gradp} is included to promote a null pressure gradient at the Windkessel outlets since the coupled equations between the 3D and 0D systems assumed the pressure to be constant at the interface. In all cases, $|| \cdot  ||$ represents the $L_2$ norm of the quantities. We make all equations dimensionless for the loss function to obtain normalized predictions of velocity and pressure, which improves the training procedure. We use the following normalization of the physical quantities:
\begin{eqnarray*}
\vec{x} \longrightarrow \vec{x}/L \\
\vec{u} \longrightarrow \vec{u}/U \\
t  \longrightarrow t/(L/U) \\
p  \longrightarrow p/(\rho U^2)
\end{eqnarray*}

where $U$ and $L$ are a characteristic velocity and length scale of the system, respectively. Consequently, the Reynolds number $Re$, defined as $Re = \frac{\rho U L }{\mu}$, appears in the Laplacian term of Equation \eqref{ec:loss_terms}. The Windkessel system is also dimensionally reduced as:
\begin{eqnarray*}
\pi  \longrightarrow \pi/(\rho U^2) \\
Q  \longrightarrow Q/(UL^2) \\
R_{d,p}  \longrightarrow R_{d,p}/(\rho U/L^2) \\
C  \longrightarrow C/(L^3/\rho U^2) 
\end{eqnarray*}

To fix the blood pressure level of the coupled system, we also give to the inverse problem the average pressure curve (obtained from the reference solution) defined as:
\begin{equation}
\label{ec:data_term}
    \Bar{p} = \sum_{k=0}^K \frac{1}{area(\Gamma_k)} \int_{\Gamma_k} p dS.
\end{equation}
Furthermore, we interpolate the latter curve in time to match the MRI measurements timestep, so both measurements set (velocity and mean pressure) are consistent in time. Note that for the steady problem, this curve consists of a single value representing the systolic pressure of the artery. In a clinical scenario, this mean pressure curve could be approximated as the pressure measured using a sphygmomanometer or an oscillometric blood pressure monitor, generally placed at the patient's upper arm or wrist \cite{saul1996non, cheng2010estimation,forouzanfar2015oscillometric}. 

Concerning the collocation points in which the term $\mathcal{L}_{NS}$ is enforced, we use $403,223$ points corresponding to a tetrahedral generated mesh for later usage. For each point, we generate a random time within the cardiac cycle as input to the network in the transient problem. Also, training by mini-batch is performed using only around $1\%$ of these points at every iteration. 

Furthermore, the coupling between the models requires the computation of a boundary integral (the velocity flow $Q_k$ at $\Gamma_k$). For that reason, all the points at the outlets have to be evaluated at the same time. To optimize the computational resources, we evaluate the velocity of every outlet point only at the measurement times (i.e., $22$ time intervals within the cardiac cycle) and at the end of every epoch. With this set of points, the flow is estimated using a finite element quadrature and interpolated in time using a cubic interpolator for later evaluation at any time within the cardiac cycle.

The transient problem presents its unique challenges compared to the steady problem. For this reason, we take into account some additional considerations that were not necessary for the steady problem:

\begin{itemize}
\item[\textit{i)}] We use a vector potential representation of the velocity $\vec{u}$ to hard-impose mass conservation of the solution. Consequently, the velocity is assumed to be originated from the curl of a vector potential $\Phi (\vec{x},t)$ \cite{janich2001vector} as:
\begin{equation}
\vec{u}(\vec{x},t) = \nabla \times \Phi (\vec{x},t).
\label{ec:phi_def}
\end{equation}
With this definition, the velocity field $vec{u}$ is always divergence free. We find that this change of variable improves the quality of the results and convergence of the method at the expense increasing the computational time.
\item[\textit{ii)}] 
We tune each weight of the total loss function automatically following the algorithm presented in \cite{wang2021understanding}, using the $L_1$ norm of the gradient of the physics loss. We perform the update every 10 epochs. The new values of the weights are exponentially averaged with their respective history following the rule: $$ \lambda_{new,i}= (1-\alpha) \cdot \lambda_{old,i} + \alpha \cdot \frac{|| \nabla \mathscr{L}_{phys} ||_{1}}{ \overline{\nabla \mathscr{L}_{i}} }  \quad \quad \forall i = 1,2, ..., N_L $$ 
being $N_L$ the number of loss terms. In all cases, we set $\alpha$ to $0.1$.

\item[\textit{iii)}] We divide the training into two stages. First, we randomly initialize the Windkessel model parameters $R_{p,k},R_{d,k},C_k$ and fix them. Then, we train only the network parameters $\mathbb{W},\vec{b}$. In a second stage, we allow the optimizer to also change the Windkensell model parameters. We found that splitting the training in this way leads to better results since the network had time first to adapt the representation of physical velocities and pressures and then to estimate the Windkessel parameters.
\item[\textit{iv)}] We compute the compliance parameter $C$ from the distal resistance $R_d$ and an estimation of the decay time $R_d \cdot C$. This time was obtained at the beginning of the training from the mean pressure curve used as an extra measurement. Furthermore, the decay time was computed by fitting an exponential curve from the valve closure time and the end of the diastole, an interval at which the Windkessel model assumes a decay behavior modulated by $\sim \exp(-t/R_dC)$. The decay time was assumed constant for all outlets at the aorta.
\end{itemize}

Finally, we obtained the weights and biases by minimizing the total loss function as:
\begin{equation}
    \mathbb{W}, \vec{b} = \arg \min_{W,b} \ \mathscr{L}_{tot}(W,b) .
\end{equation}

\begin{table}[h] \centering
\footnotesize
\newcolumntype{C}{>{\centering\arraybackslash}X}
\sisetup{table-format=1, table-number-alignment=center}
\begin{tabular}{cllllllll}
\toprule
\addlinespace
 & &  \multicolumn{3}{c}{\makecell{Steady Problem}} & \multicolumn{3}{c}{\makecell{Transient Problem}}  & 
\tabularnewline
\cmidrule[\lightrulewidth](lr){1-9} \addlinespace[1ex]
Estimation  & &  & {$ \vec{u}$ , $p$ , $ \big \{ R_{tot} \big \}^{5}_{k=1}$} & &   {$ \vec{u}$ , $p$ , $ \big \{ R_{p}, R_d , C_d \big \}^{5}_{k=1}$ }
\tabularnewline
\addlinespace
\bottomrule
\end{tabular}
\caption{Different quantities to estimate in the steady and transient problems.}
\label{tab:num_exp}
\end{table}

Figure \ref{fig:NN} shows a schematic of the final architecture used for the transient problem. The minimization was performed using the ADAM optimizer \cite{kingma2014adam} with the \emph{swish} activation function \cite{ramachandran2017searching}. We perform the estimation of the velocity and pressure fields and the estimation of the Windkessel parameters simultaneously. The Winkessel parameters start from a random seed between a predefined confidence interval between the half and the double of the respective reference value and then evolve with the optimization. Five independent realizations are performed in which the network weights are randomly initialized in each one of these. All cases are summarized in Table \ref{tab:num_exp}.

The training of all the experiments were done in an NVIDIA RTX-A6000. The codes were written in Python using the PyTorch library \cite{NEURIPS2019_9015} and are available on GitHub at \texttt{https://github.com/yeyemedicen/PINNs-WK-MRI}. All used hyperparameters are reported for both stationary and transient problems in Table \ref{tab:hyperparameters}.

\begin{table}[htbp] \centering
\footnotesize
\newcolumntype{C}{>{\centering\arraybackslash}X}
\sisetup{table-format=1, table-number-alignment=center}
\begin{tabular}{cS[table-format=1]*{7}{S}}
\toprule
\addlinespace
 {Hyperparameter} & &  \multicolumn{3}{c}{\makecell{Steady Problem}} & & \multicolumn{3}{c}{\makecell{Transient Problem}}
\tabularnewline
\cmidrule[\lightrulewidth](lr){1-9} \addlinespace[1ex]
Batchsize & &  &  {$1580$} & & & & {$3000$}
\tabularnewline
Learning Rate  & &  & {$10^{-3}$} & & & & {$10^{-3}$}
\tabularnewline
Learning Rate of Parameters  & &  & {$10^{-2}$} & & & & {$10^{-2}$}
\tabularnewline
Learning Rate Scheduler &  &  & {Yes} & & & & {Yes}
\tabularnewline
Hidden layers &  &  & {$7$} & & & & {$7$}
\tabularnewline
No. neurons per layer &  &  & {$220$} & & & & {$220$}
\tabularnewline
Hidden layers ($\pi$) &  &  & {$-$} & & & & {$6$}
\tabularnewline
No. neurons per layer ($\pi$) &  &  & {$-$} & & & & {$10$}
\tabularnewline
Activation Function &  &  & {swish} & & & & {swish}
\tabularnewline
Epochs &  &  & {$1250$} & & & & {$120 + 1250$ }
\tabularnewline
Potential Vector Representation &  &  & {No} & & & & {Yes}
\tabularnewline
Normalized Length ($L$) &  &  & {$1$ cm} & & & & {$0.5$ cm}
\tabularnewline
Normalized Velocity ($U$) &  &  & {$5000$ cm/s} & & & & {$120$ cm/s}
\tabularnewline
Loss weights &  &  & {Manual} & & & & {Automatic}
\tabularnewline
\cmidrule[\lightrulewidth](lr){2-8}
 & & {$\lambda_{phys}$}   & {1.5} & & & & {$-$}
 \tabularnewline
 & & {$\lambda_{data,u}$}   & {1.0} & & & & {$-$}
 \tabularnewline
 & & {$\lambda_{data,p}$}   & {1.0} & & & & {$-$}
 \tabularnewline
 & & {$\lambda_{BC}$}   & {6.0} & & & & {$-$}
\tabularnewline
 & & {$\lambda_{windk}$}   & {1.0} & & & & {$-$}
\tabularnewline
\cmidrule[\lightrulewidth](lr){2-8}
\addlinespace
\bottomrule
\end{tabular}
\caption{Hyperparameters used for training the PINNs}
\label{tab:hyperparameters}
\end{table}

The normalized velocity and length values were tuned by hand, keeping in mind that the normalized quantities entering the PINNs workflow have to be ideally not larger than the unity. In the case of the transient problem, the value of the normalized velocity had to be significantly reduced because of the lower velocities at the end of the cardiac cycle (diastole).

\subsection{The Reference Solution}
To create the reference solution, equations \eqref{eq:NS} and \eqref{eq:WK} are solved using the finite element method (FEM), discretizing in space using stabilized $\mathbb{P}1/\mathbb{P}1$ elements, and in time using a backward Euler scheme. The system was solved using a non-incremental fractional step scheme detailed in Ref \cite{garay2022parameter}. The initial conditions were set to: 
\begin{equation*}
\label{eq:initial_conditions}
\begin{cases}
\displaystyle \vec{u} =\vec{0} , \\
\pi_k = 78.8 \ \rm{mmHg} \quad \text{for} \quad k=1, \dots,K .
\end{cases}
\end{equation*}

\begin{figure}[!h]
\centering
\subfloat[]{
  \includegraphics[trim=500 250 900 500, clip, width=0.32\textwidth]{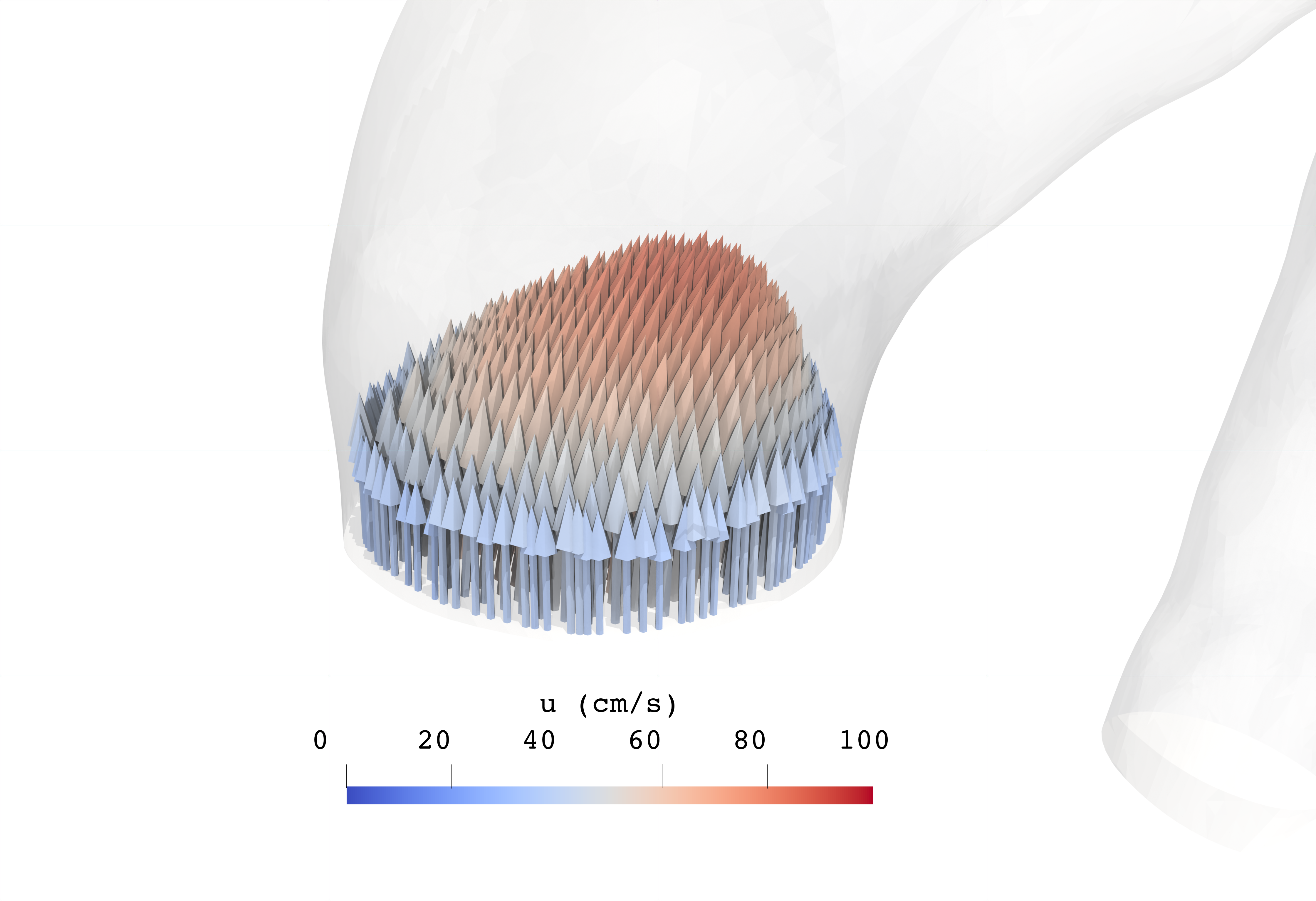} } 
\subfloat[]{
  \includegraphics[trim=0 0 0 0, clip, width=0.40\textwidth]{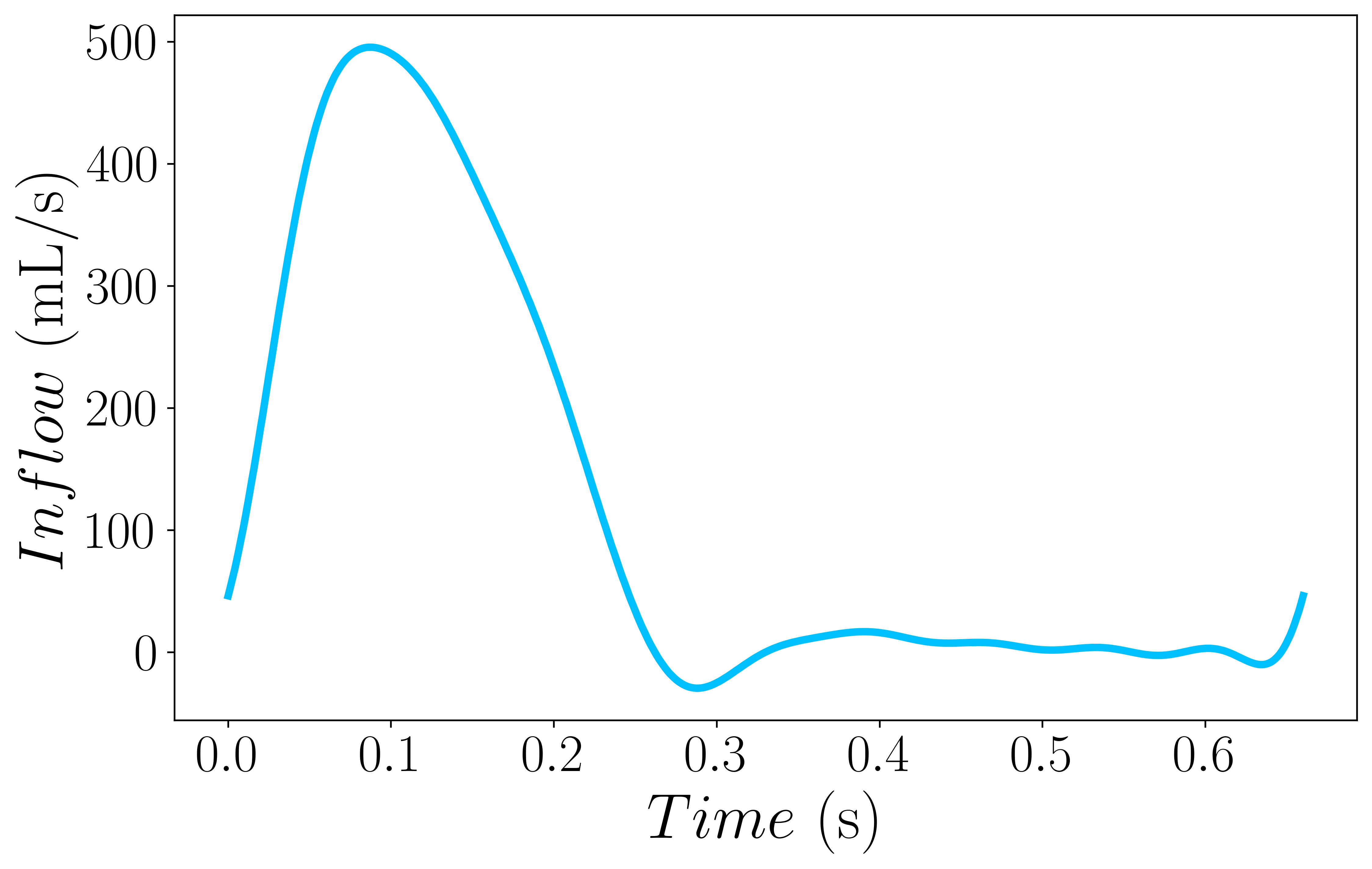} } 
  \caption{(a) Stokes-profile velocity at the inlet of the aorta. (b) Velocity flow at the inlet taken from the SimVascular dataset.}
  \label{fig:sim_inflows}
\end{figure}

The values of $\pi_k$ correspond to approximately the periodic state of the 3D-0D system taken from Ref \cite{wilson2013vascular}. Finally, we assume a velocity profile at $\Gamma_{inlet}$ as:
\begin{equation}
\vec{u}(\vec{x},t)_{inlet} = \vec{u}_{stokes}(\vec{x})  f(t) \ \vec{n},
\end{equation}
where $\vec{u}_{stokes}(\vec{x})$ is the solution of a Stokes problem to ensure a parabolic-shape solution already adapted to the domain as is shown in Figure \ref{fig:sim_inflows}(a). The function $f(t)$ represents the time-dependency of the inflow velocity and is taken in such a way that the total flow through $\Gamma_{in}$ follows the curve in Figure \ref{fig:sim_inflows}(b), obtained from the Vascular Model Repository database \cite{wilson2013vascular}. The whole boundary condition is then added into the variational form as a penalization term $A_{inlet}$ defined as:
\begin{equation}
A_{inlet} = \gamma \int_{\Gamma_{inlet}} (\vec{u} - \vec{u}_{inlet}) \cdot \vec{v}
\end{equation}
with the parameter $\gamma$ fixed in $10^5 \ \rm{gr/(cm^2 \cdot s)}$.

The Windkessel parameters were also obtained from the aforementioned repository database. For the numerical values of these constants, see Table \ref{tab:Parameters}.

\begin{table}[!h]
\footnotesize
\centering
\begin{tabular}{cS[table-format=1]*{5}{S}}
\toprule
\addlinespace
{Parameter} &  & {$\Gamma_1$} & {$\Gamma_2$} & {$\Gamma_3$} & {$\Gamma_4$} & {$\Gamma_5$} 
\tabularnewline
\cmidrule[\lightrulewidth](lr){1-7}\addlinespace[1ex]
$R_p \ (\rm{dyn \cdot s \cdot cm^{-5}})$ & & {713} & {713} & {602} & {689} & {98}\tabularnewline
\cmidrule[\lightrulewidth](lr){1-7}\addlinespace[1ex]
$R_d \ (\rm{dyn \cdot s \cdot cm^{-5}})$ &  & {12023} & {12023} & {10143} & {11609} & {1650}
\tabularnewline
\cmidrule[\lightrulewidth](lr){1-7}\addlinespace[1ex]
$C_d \ (\rm{dyn^{-1} \cdot cm^5} )$ & & {8.256e-5} & {8.256e-5} & {9.785e-5} & {8.55e-5} & {6.015e-4}\tabularnewline
\bottomrule
\end{tabular}

\caption{Three-element Windkessel parameters for every outlet.
        } \label{tab:Parameters}
\end{table}

The computational domain corresponds to a thoracic aorta with a total stream-wise length of $10.8 \ \rm{cm}$ and meshed with 558,112 tetrahedrons and 115,779 vertices. We assume blood has a density of $1.06 \ \rm{gr/cm^3}$ and a constant dynamic viscosity of $0.035$ P. The time step was set to $\tau = 0.001 \ \rm{s}$ with a total run time of $0.66 \ \rm{s}$. We also added backflow stabilization in every system outlet \cite{bazilevs2009patient}. 

\begin{figure}[!hbtp]
\centering
\subfloat[]{
  \includegraphics[trim=100 0 150 00, clip, width=0.32\textwidth]{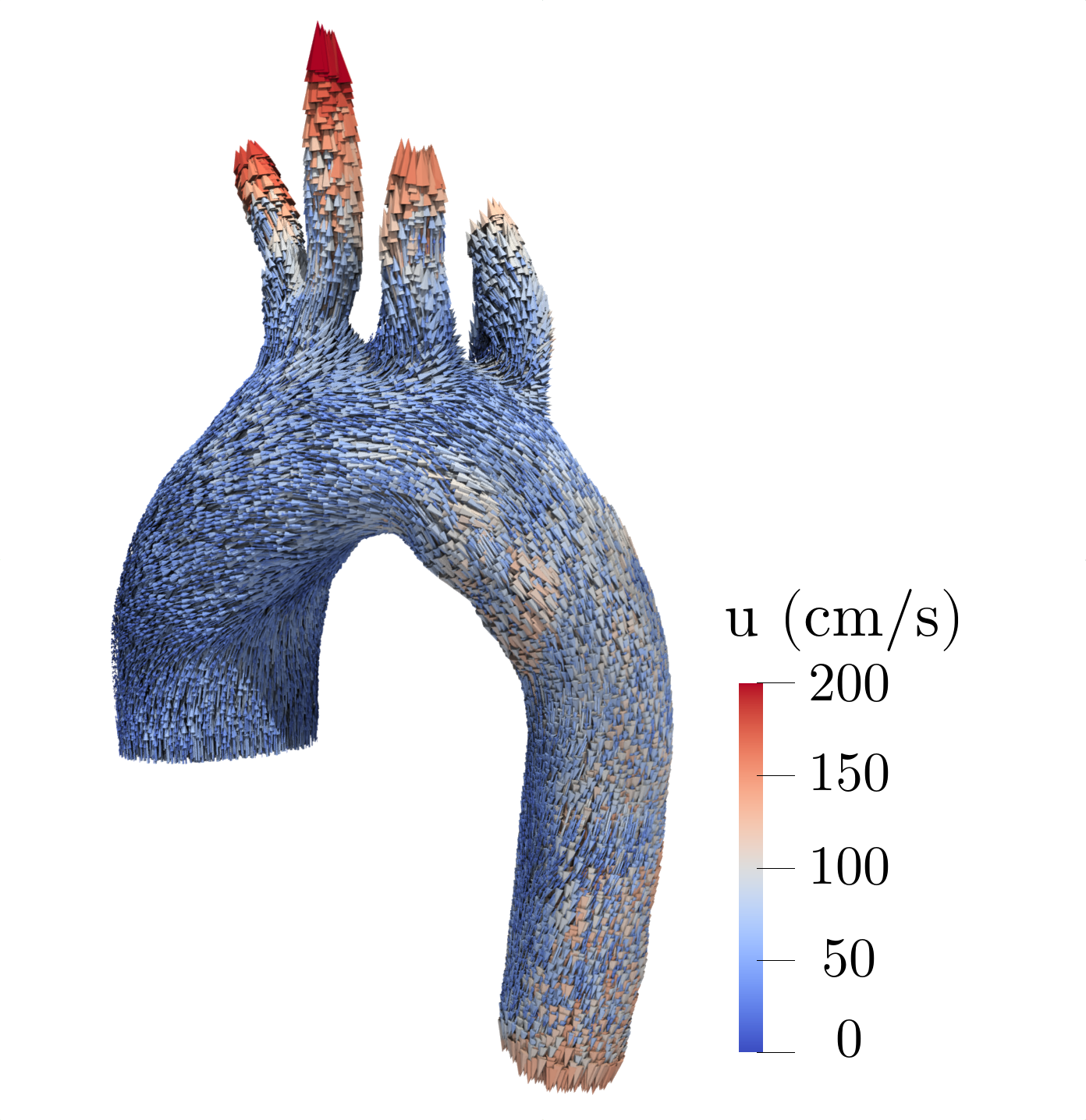} }
  \hspace{0.2cm}
  \subfloat[]{
  \includegraphics[trim=0 0 0 0, clip, width=0.28\textwidth]{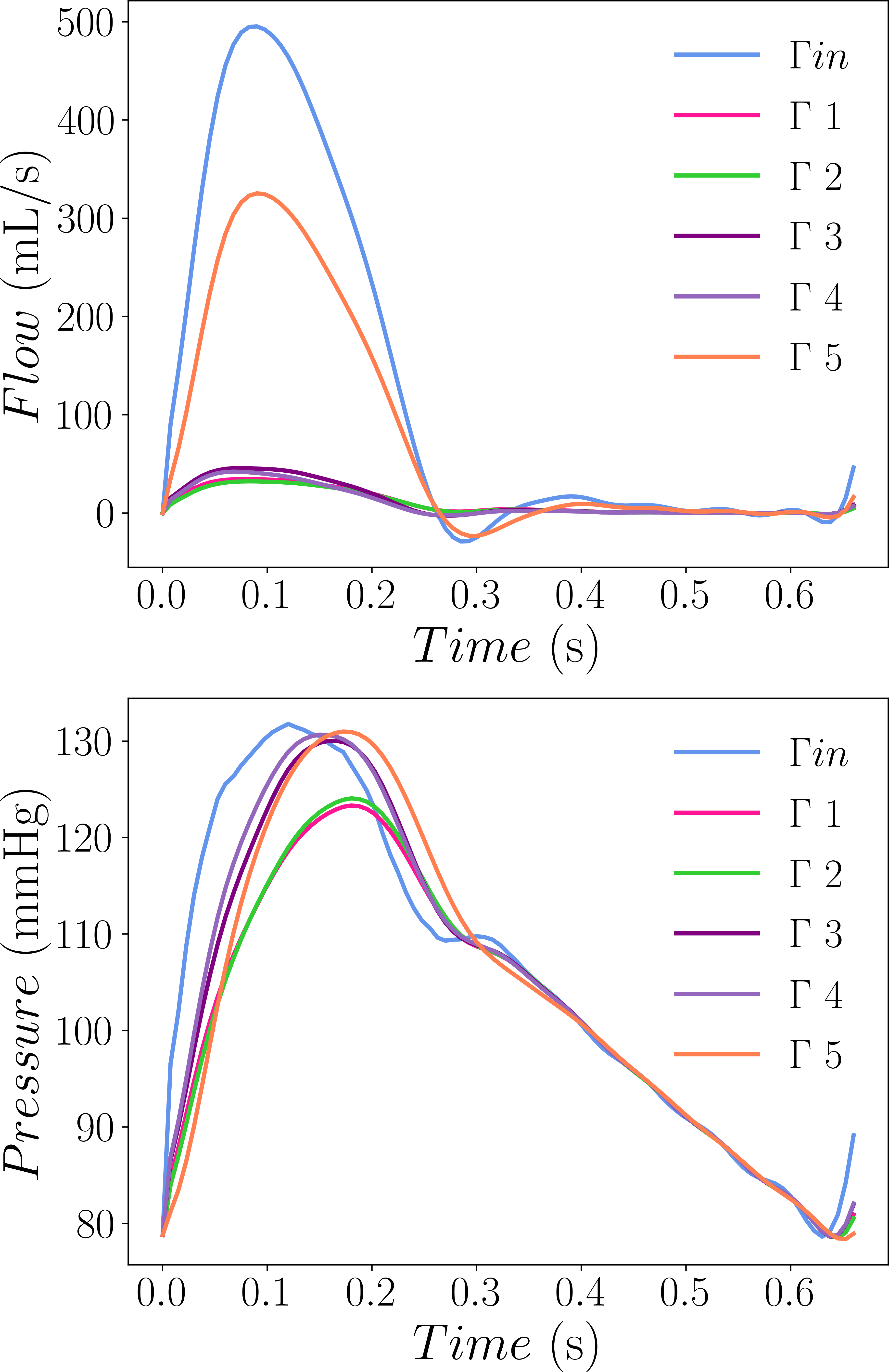} }
  \caption{Reference simulation results. The blood flow velocity at peak systole and its flow rates and average pressure curves are depicted in (a) and (b), respectively.}
\label{fig:reference_sol}
\end{figure}

Figure \ref{fig:reference_sol} shows the resulting velocity field at peak systole next to each outlet's flow rates and pressure curves.

\subsection{PC-MRI measurement generation}
\begin{figure}[!hbtp]
\centering
\subfloat[]{
  \includegraphics[trim=0 -200 0 0, clip, width=0.42\textwidth]{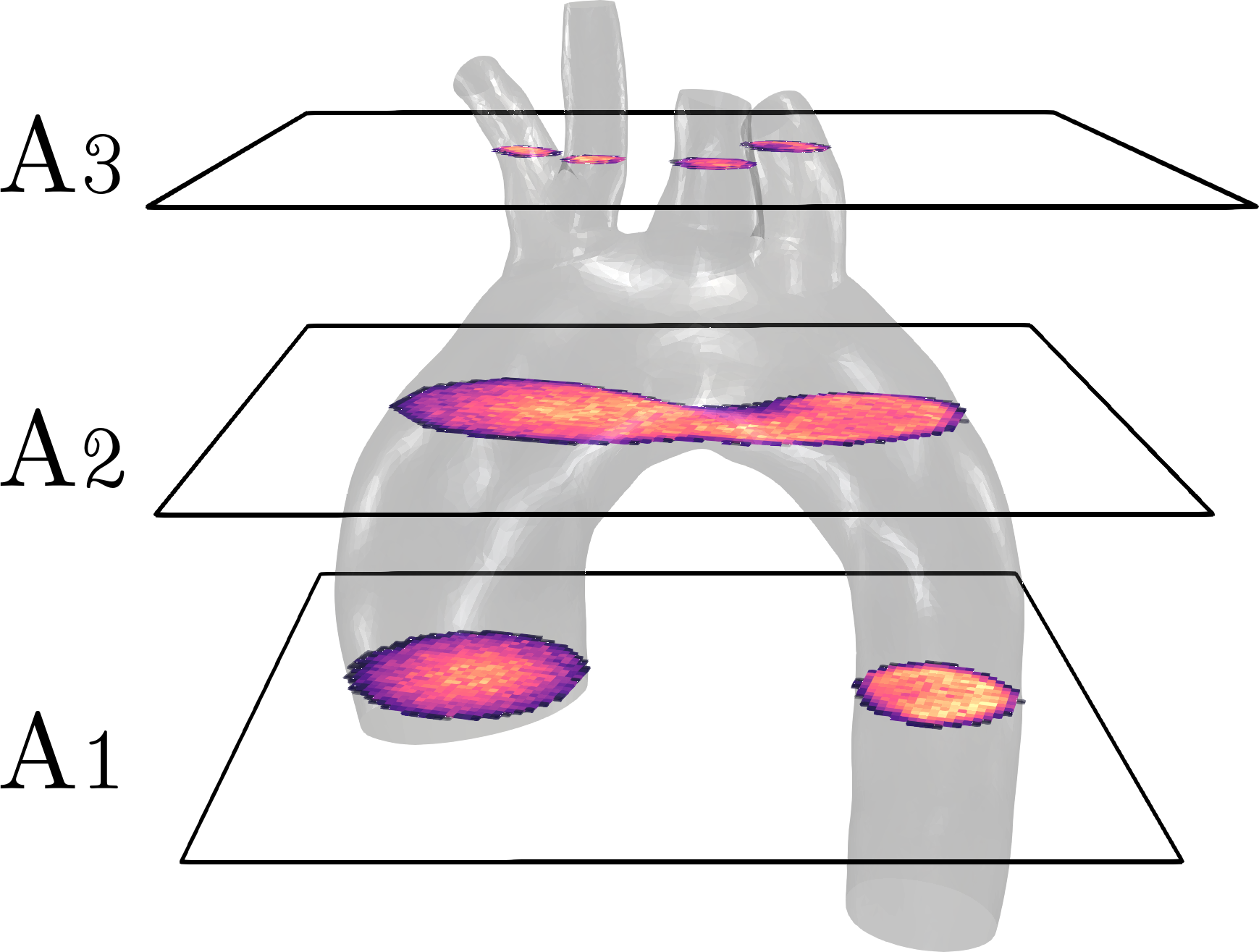} }
  \hspace{0.5cm}
\subfloat[]{
  \includegraphics[trim=0 0 0 0, clip, width=0.3\textwidth]{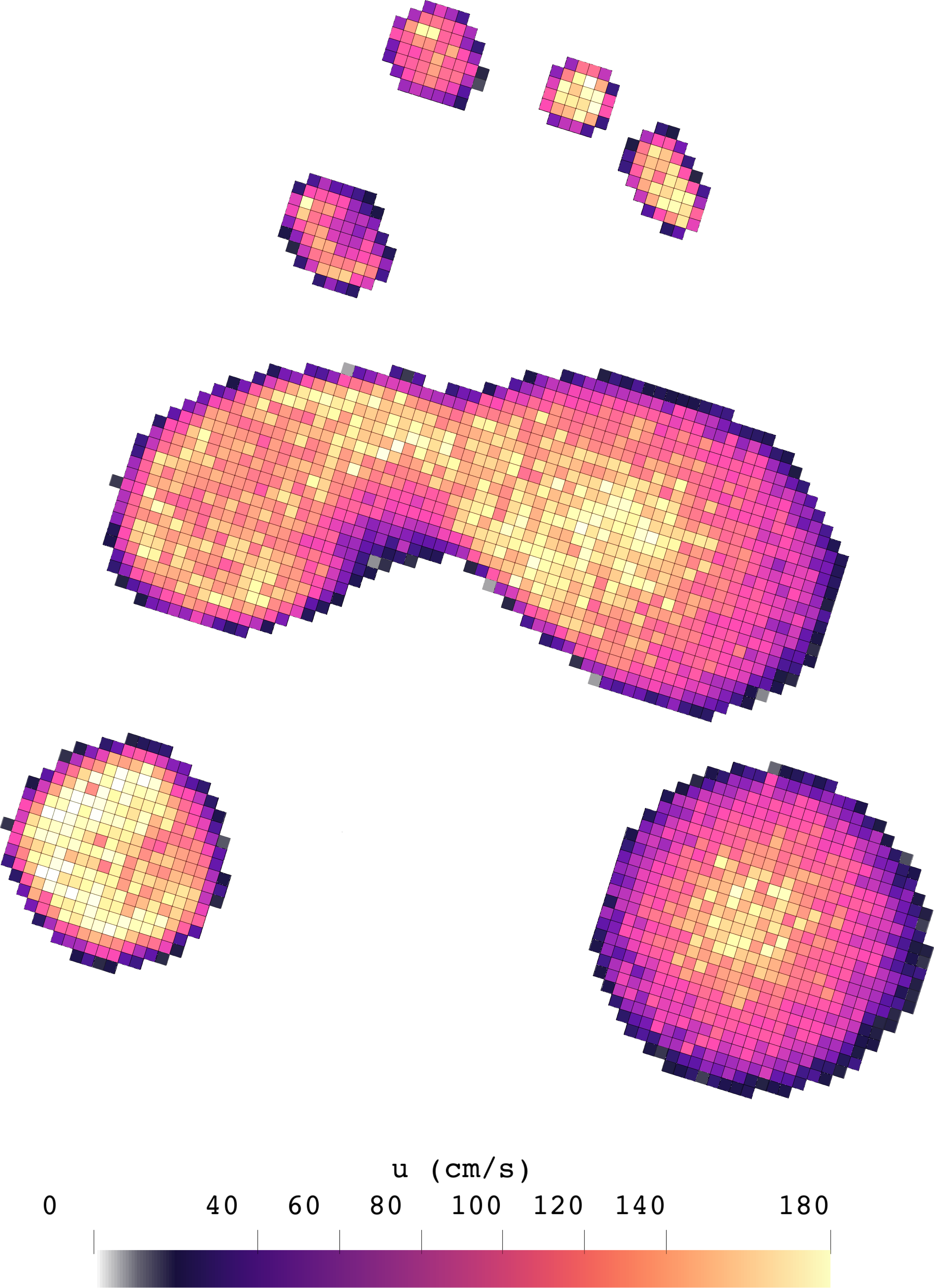} }
  \caption{Synthetic PC-MRI measurements obtained from the simulation. The axial planes taken from the aorta are shown in (a), while the resulting PC-MRI velocity measurements are shown in from frontal view in (b).}
\label{fig:mri_meas}
\end{figure}

From the reference solution, we simulate a 2D phase-contrast magnetic resonance acquisition located at three axial planes covering the whole domain, as shown in Figure \ref{fig:mri_meas}(a). Thus, the velocity field is first interpolated into a voxel-like slice mesh with a resolution of $1.0 \times 1.0 \ \rm{mm^2}$  and downsampled in time from $1 \ ms$ to $30 \ ms$, following the typical timestep of a clinical acquisition. After that, two different complex magnetization vectors are defined as:
\begin{eqnarray}
M^u_{meas} & = & M_0 \ \exp(i\phi_0 + i \pi u_{MRI}/venc) , \\
M^0_{meas} & = & M_0 \ \exp(i\phi_0) ,
\end{eqnarray}
where $M^u_{meas}$ is the velocity-encoded magnetization and $M^0_{meas}$ is an extra measurement usually done to capture the reference phase $\phi_0$. The interpolated velocity is represented by $u_{MRI}$ while $M_0$ is assumed constant within the vessel's lumen.
Moreover, Gaussian noise is added to the magnetization components, producing a fixed signal-to-noise ratio of $18$ dB. Consequently, a non-Gaussian noise distribution is induced in the reconstructed velocity, as is generally observed in PC-MRI measurements \cite{IRARRAZAVAL2019250}. The final velocity field is then reconstructed using the following equation: 
\begin{equation}
\label{ec:pc_mri}
 \vec{u}_{meas} = venc \ \frac{\angle \big ( M^u_{meas} / M^0_{meas} \big ) }{\pi} ,
 \end{equation}
 with the symbol $\angle$ representing the angle of the complex quantity measured from $-\pi$ to $\pi$. The velocity encoding parameter ($venc$) is chosen as the $120 \%$ of the maximum velocity to ensure aliased-free measurements. Figure \ref{fig:mri_meas}(b) shows the resulting slice measurements for the stationary problem. We perform a similar procedure for the transient flow solution, resulting in $22$ phases along the cardiac cycle. 

\section{Results}
\label{sec:results}
\subsection{Steady problem}
In this section, we present the results obtained for the stationary problem. For each experiment, statistics are presented of 5 independent experiment realizations.

\begin{figure}[h]
\centering
\includegraphics[width=1.0\textwidth]{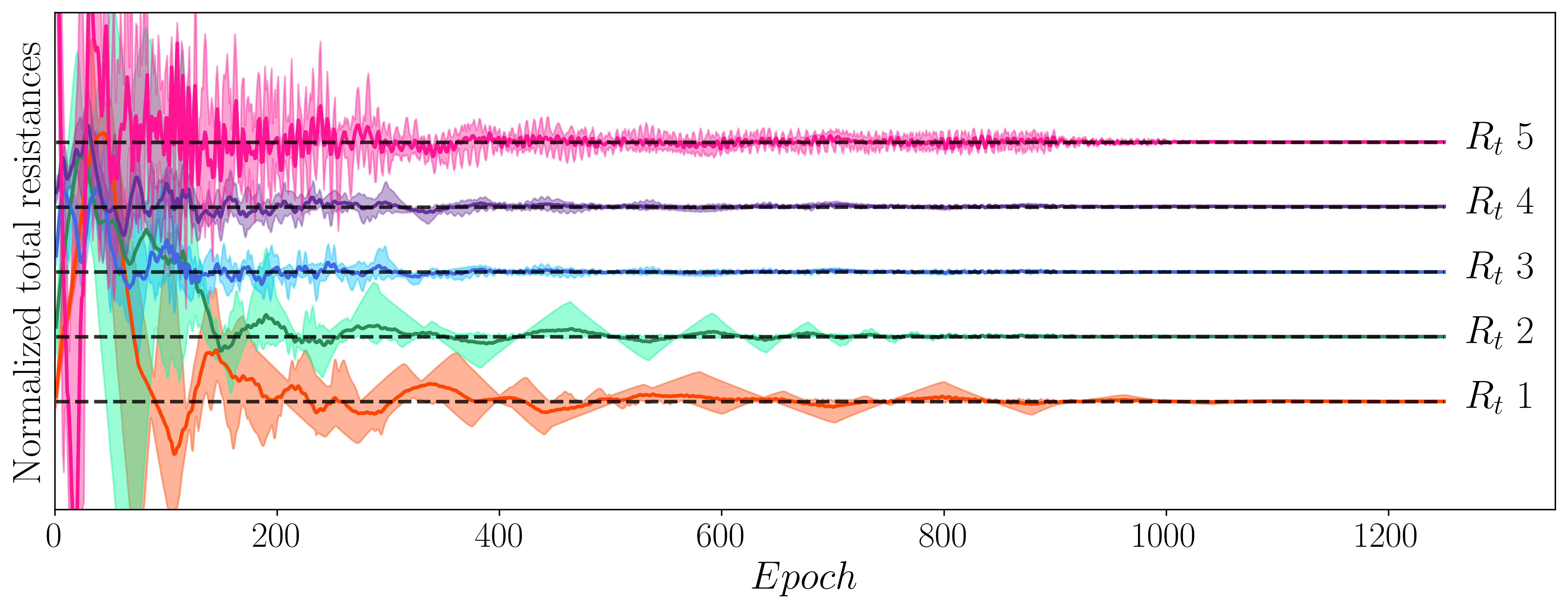}
\caption{Evolution of Parameter estimation statistics for each total resistances of the aorta in steady flow regime. All values were divided by their reference value and split vertically for better visualization. Dashed black lines represent the exact solution, and continuous color lines represent the estimation mean. The colored area surrounding the curves is determined by the minimum and maximum value found along the multiple realizations of the experiment}
\label{fig:steady_params}
\end{figure}

Figure \ref{fig:steady_params} shows the evolution of the estimated total resistance by every training epoch and Windkessel outlet. We normalize all values to ease the visualization of the parameter evolution. It can be seen that all parameters stabilize rapidly to their reference value, yet oscillations are present; they reduce at the last stages of the training. Also, not all parameters show the same convergence rate since, for instance, the outlet number $3$ takes around one-third of the total training epochs to converge to its reference value, while the outlet number $5$ needs to approximately double the iterations to have a same-accurate result.

\begin{table}[htbp] \centering
\footnotesize
\newcolumntype{C}{>{\centering\arraybackslash}X}
\sisetup{table-format=1, table-number-alignment=center}
\begin{tabular}{cS[table-format=1]*{10}{S}}
\toprule
\addlinespace
 &  \multicolumn{10}{c}{\makecell{Steady Problem}} &   \\
  \cmidrule(lr){2-12}
 & \multicolumn{3}{c}{\makecell{$ Q \ (\rm{mL/s})$}} & & \multicolumn{3}{c}{\makecell{$P \ (\rm{mmHg})$}} &  &  \multicolumn{3}{c}{\makecell{$R_{tot} \  (10^3 \ \rm{dyn \cdot s \cdot cm^{-5}})$}}  \\
  \cmidrule(lr){2-4} \cmidrule(lr){6-8} \cmidrule(lr){10-12}
{Bnd.} & {Ref.} &{Mean}& {Std} & & {Ref} & {Mean } &{Std} & & {Ref} & {Mean} &{ Std }  
\tabularnewline
\cmidrule[\lightrulewidth](lr){1-12} 
\addlinespace[1ex]
$\Gamma_{in}$ 	  &  {-298.88} & {-297.16} & {2.07} & & {248.39} & {248.92} & {0.63}  & &  {$-$} & {$-$} & {$-$}  \tabularnewline
$\Gamma_1$ & {25.82}  & {25.58} & {0.10} & & {247.96} & {246.93} & {1.63}  & &  {12.74} & {12.86} & {0.14} \tabularnewline
$\Gamma_2$	  &  {25.69}  & {25.64} & {0.13} & & {247.90} & {247.34} & {0.80}  & &  {12.74} & {12.89} & {0.06}  \tabularnewline
$\Gamma_3$	  &  {30.56}  & {30.71} & {0.27} & & {248.20} & {247.98} & {0.66}  & &  {10.75} & {10.78} & {0.08}  \tabularnewline
$\Gamma_4$	  &  {26.65}  & {26.08} & {0.12} & & {248.22} & {249.08} & {0.66}  & &  {12.30} & {12.71} & {0.10}  \tabularnewline
$\Gamma_5$	  &  {188.85} & {185.97} & {0.84} & & {248.15} & {248.30} & {0.66}  & &  {1.74} & {1.78} & {0.04}  \tabularnewline
\addlinespace
\bottomrule
\end{tabular}
\caption{Flow and mean pressure for each outlet of the aorta. The mean pressures were computed as $ P_i = \frac{1}{Area_i}\int_{\Gamma_i} p ds$, where $i$ is the corresponding outlet number. The mean and standard deviation values reported are computed from the results of every seed.}
\label{tab:steady_results}
\end{table}

\begin{figure}[!h]
\centering
\subfloat[Reference Velocity]{
  \includegraphics[trim=800 0 900 0, clip, width=0.38\textwidth]{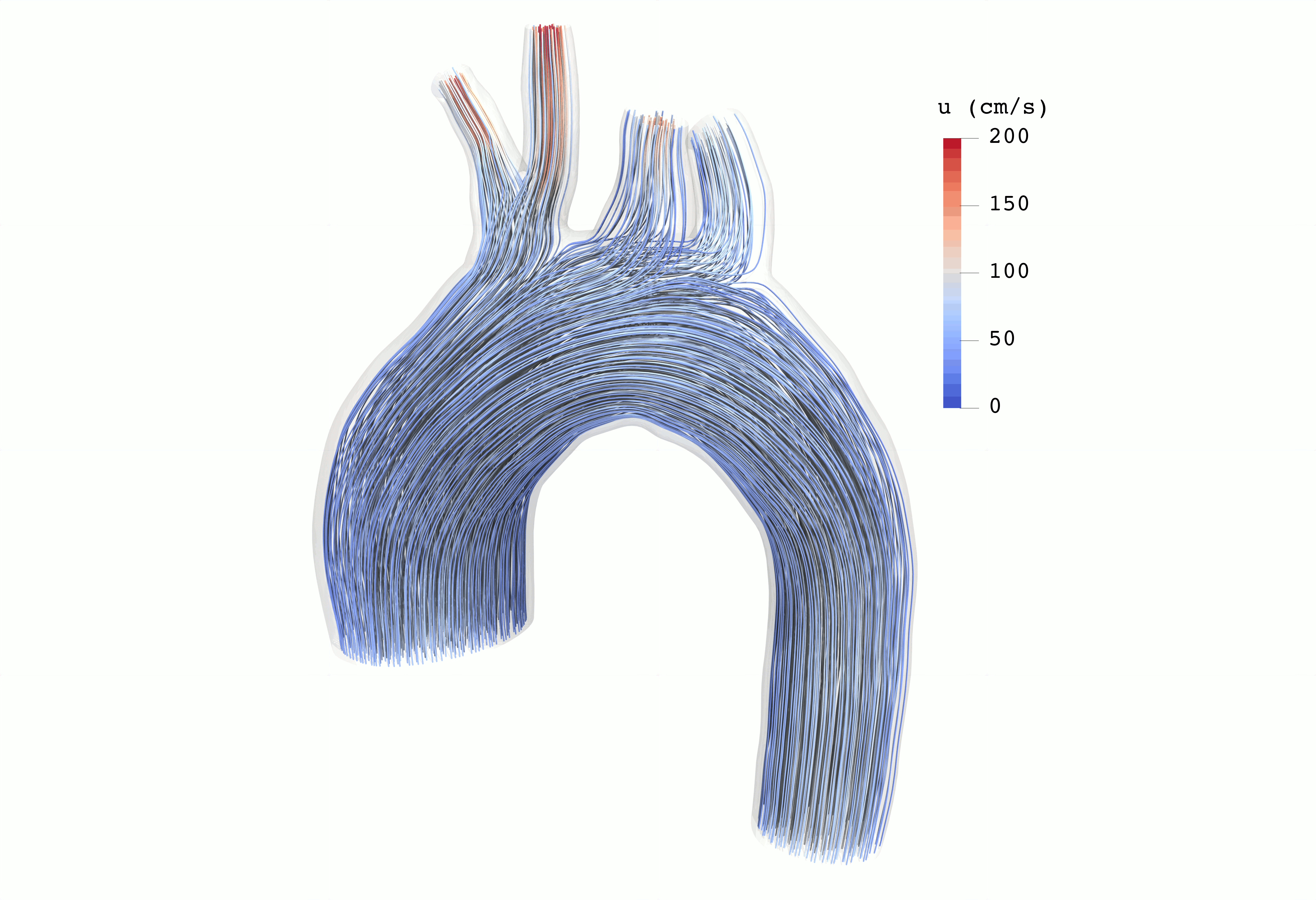} } 
\subfloat[Mean Estimated Velocity]{
  \includegraphics[trim=800 0 900 0, clip, width=0.38\textwidth]{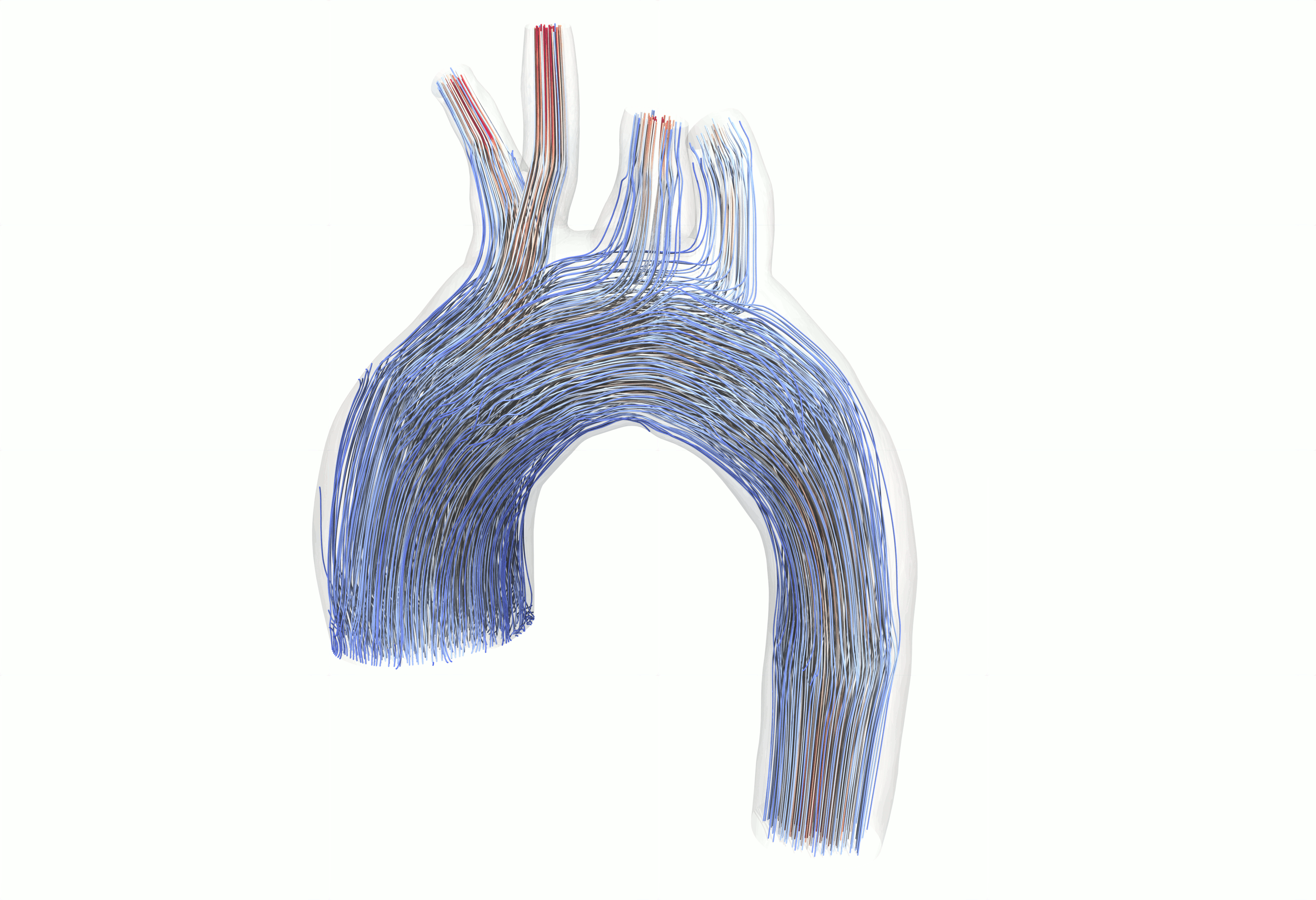} }
  \caption{Velocity streamlines when the steady-state flow regime is imposed for (a) the reference solution and (b) the mean estimated velocity represented by the PINNs}
  \label{fig:streamlines_steady}
\end{figure}

Table \ref{tab:steady_results} shows the final values of the estimation, the mean estimated flow, and pressure at every outlet. These quantities were computed from the mean of the velocities and pressures by every experiment realization. We have found that the PINNs represent the solution at the inlet remarkably well, even though no information was given at that boundary. Also, the estimated parameters match their corresponding reference values well despite the lack of volumetric data and the artifacts introduced to the slice measurements due to the added noise and spatiotemporal interpolation. Finally, Figure \ref{fig:streamlines_steady} shows the velocity streamlines computed from the inlet nodes of the mesh of both reference and mean estimation velocities. This figure shows that the PINNs' solution is similar to the reference velocity while in the vessel's lumen and moderately fails for the nodes closer to the walls.

\subsection{Transient problem}
\label{sec:transient_problem_results}

\begin{figure}[h]
\centering
\includegraphics[trim=0 0 0 0, clip, width=0.9\textwidth]{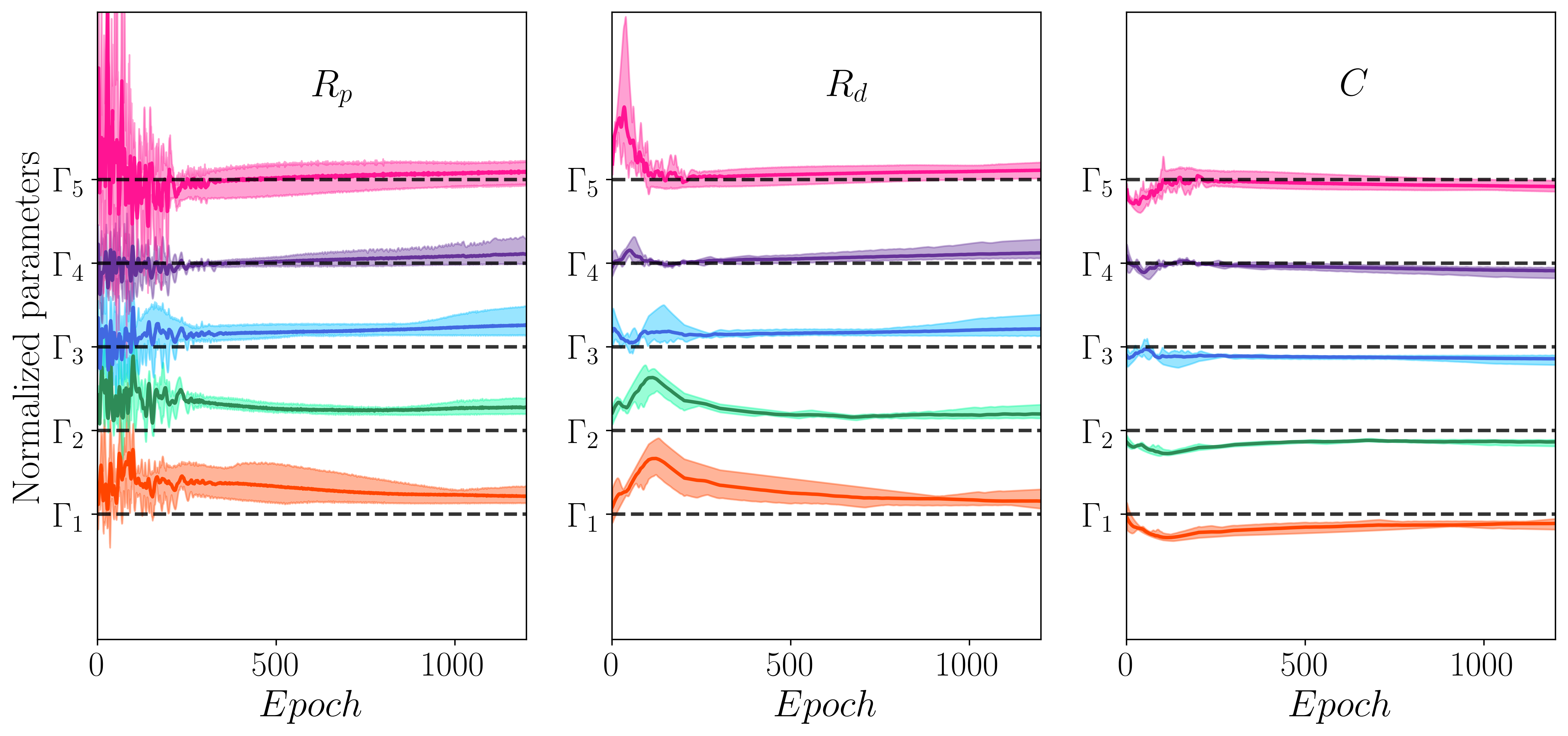}
\caption{Evolution of the parameter estimation statistics for each Windkessel outlet of the aorta in transient flow regime. All values were divided by their reference value and split vertically for better visualization. Dashed black lines represent the exact solution, and continuous color lines represent the estimation mean. The colored area surrounding the curves is determined by the minimum and maximum values found along the multiple realizations of the experiment.}
\label{fig:transient_params}
\end{figure}

Figure \ref{fig:transient_params} shows the parameter estimation evolution. As before, we normalize the parameters by their reference values and split them in the vertical axis to ease the visualization. From the plots, we can see that proximal resistances ($R_p$) show significantly more oscillation during the first epochs of the training than the distal resistances parameters ($R_d$), though both show relatively fast convergence to their respective reference values within the half of the training. However, some outlets ($\Gamma_3$ and $\Gamma_4$) present a growing standard deviation in the last epochs of the training for the estimated resistance parameters. Table \ref{tab:transient_results} reports the final values with their respective statistics.

\begin{table}[!h] \centering
\footnotesize
\newcolumntype{C}{>{\centering\arraybackslash}X}
\sisetup{table-format=1, table-number-alignment=center}
\begin{tabular}{cS[table-format=1]*{10}{S}}
\toprule
\addlinespace
 &  \multicolumn{10}{c}{\makecell{Transient Problem}} &   \\
  \cmidrule(lr){2-12}
 & \multicolumn{3}{c}{\makecell{$R_{p} \  (10^3 \ \rm{dyn \cdot s \cdot cm^{-5}})$}} & & \multicolumn{3}{c}{\makecell{$R_{d} \  (10^3 \ \rm{dyn \cdot s \cdot \rm{cm^{-5}}})$}} &  &  \multicolumn{3}{c}{\makecell{$C_{d} \  (10^{-5} \ \rm{dyn^{-1} \cdot cm^5} )$}}  \\
  \cmidrule(lr){2-4} \cmidrule(lr){6-8} \cmidrule(lr){10-12}
{Bnd.} & {Ref.} &{Mean}& {Std} & & {Ref} & {Mean } &{Std} & & {Ref} & {Mean} &{ Std }  
\tabularnewline
\cmidrule[\lightrulewidth](lr){1-12} 
\addlinespace[1ex]
$\Gamma_1$ & {0.71}  & {1.02} & {0.10} & & {12.02} & {15.76} & {1.87}  & & {8.26} & {6.42} & {0.70} \tabularnewline
$\Gamma_2$	  &  {0.71}  & {1.10} & {0.12} & & {12.02} & {16.75} & {1.58} & & {8.26} & {6.01} & {0.52}  \tabularnewline
$\Gamma_3$	  &  {0.60}  & {0.91} & {0.15} & & {10.14} & {14.51} & {1.88} & & {9.78} & {6.99} & {0.82}  \tabularnewline
$\Gamma_4$	  &  {0.69}  & {0.84} & {0.14} & & {11.61} & {14.46} & {1.94} & & {8.55} & {7.01} & {0.80}  \tabularnewline
$\Gamma_5$	  &  {0.10} & {0.12} & {0.02} & & {1.65} & {2.02} & {0.22}  & & {60.15} & {50.18} & {5.51}  \tabularnewline
\addlinespace
\bottomrule
\end{tabular}
\caption{Parameter estimation obtained on the transient-flow regime.}
\label{tab:transient_results}
\end{table}

\begin{figure}[!h]
\centering
\includegraphics[trim=0 0 0 150, clip, width=0.76\textwidth]{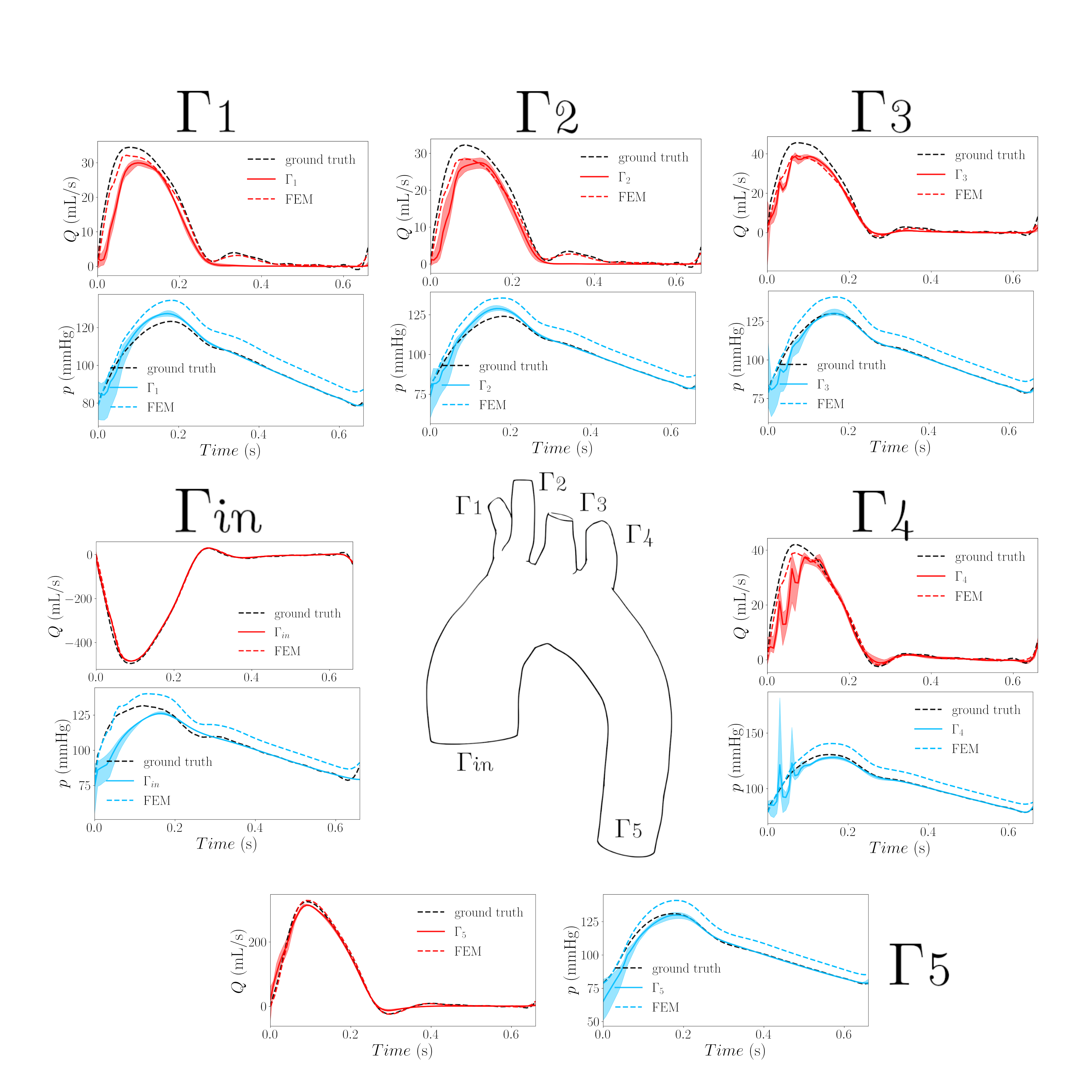}
\caption{Flows (red) and mean pressure (sky blue) curves at every aortic outlet over time. Continuous lines represent mean estimation, while the area plot is defined by the minimum and maximum value encountered in every experiment realization. Black dashed lines correspond to the respective reference value, while colored dashed lines correspond to the curves obtained from the post-process FEM simulation.}
\label{fig:transient_flows}
\end{figure}

Figure \ref{fig:transient_flows} shows the obtained flows and mean pressure curves. From these curves, a general flow underestimation is shown in the upper outlets of the aorta within the systole, while the inlet and descending aorta ($\Gamma_{in}$ and $\Gamma_5$) were remarkably well captured in the whole cardiac cycle. Conversely, mean pressure curves computed as: $$ P_{i} = \frac{1}{area(\Gamma_i)} \int_{\Gamma_i} p \  dS , $$ 
tend to be overestimated in the upper aorta and underestimated in the downward portion of the domain. Moreover, an oscillating behavior was found in the flows and pressure in the outlets $\Gamma_3$ and $\Gamma_4$, outlets where the standard deviation of the estimated parameters is the most. A FEM simulation was also performed after the training, using the estimated parameters and inlet velocity profile found in the experiment, which present the median error among all realizations. This solution could be seen as a post-process step of the PINNs' estimation and generally presents reduced oscillations compared with the raw estimated velocity and pressure fields. The obtained curves of the post-process FEM simulation are depicted in colored dashed lines.

\begin{figure}[h!]
\centering
\subfloat[\footnotesize{Reference ($t=0.12$ s)}]{
  \includegraphics[trim=900 20 900 20, clip, width=0.3\textwidth]{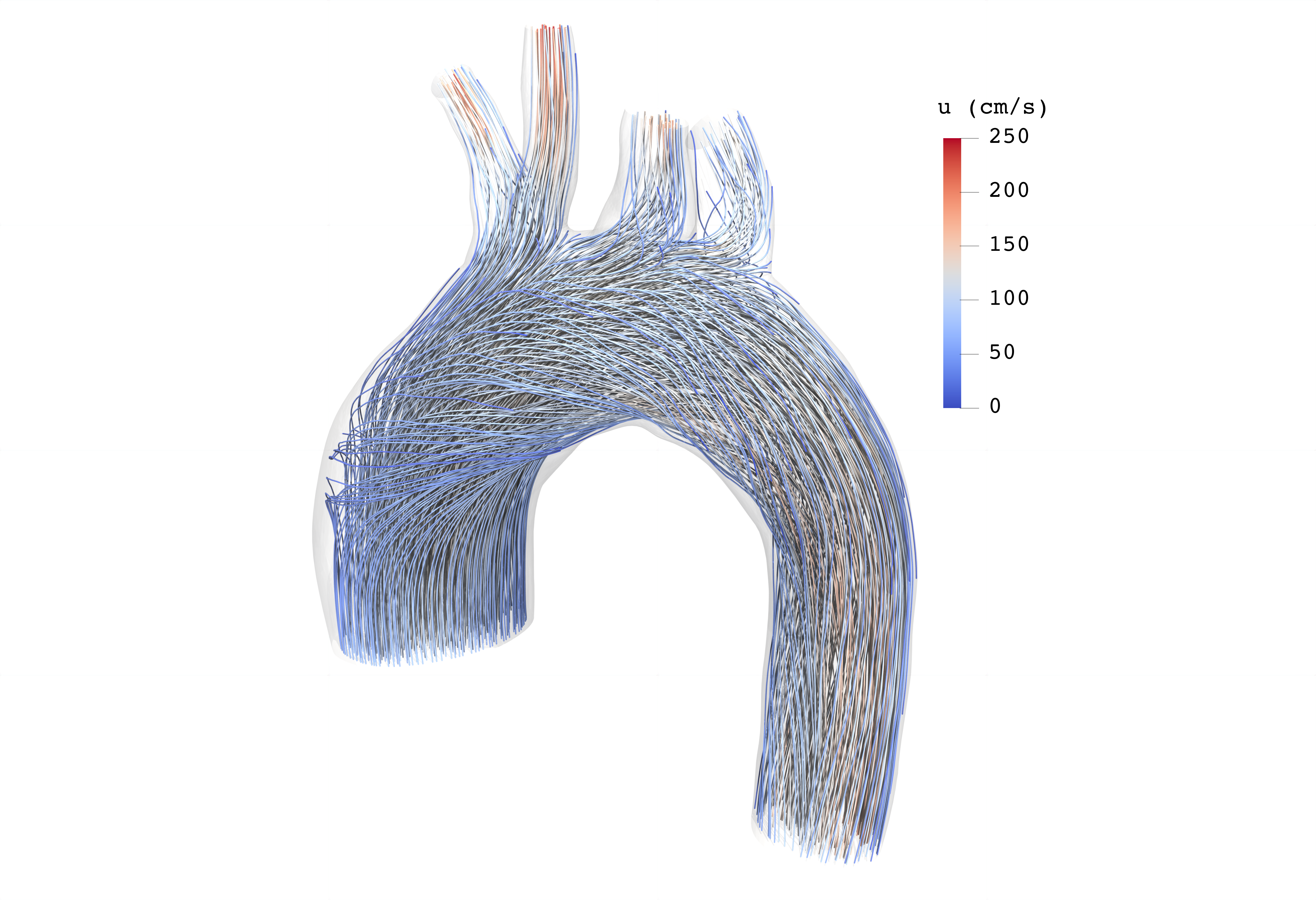} } 
\subfloat[\footnotesize{PINNs ($t=0.12$ s)}]{
  \includegraphics[trim=1000 20 900 20, clip, width=0.29\textwidth]{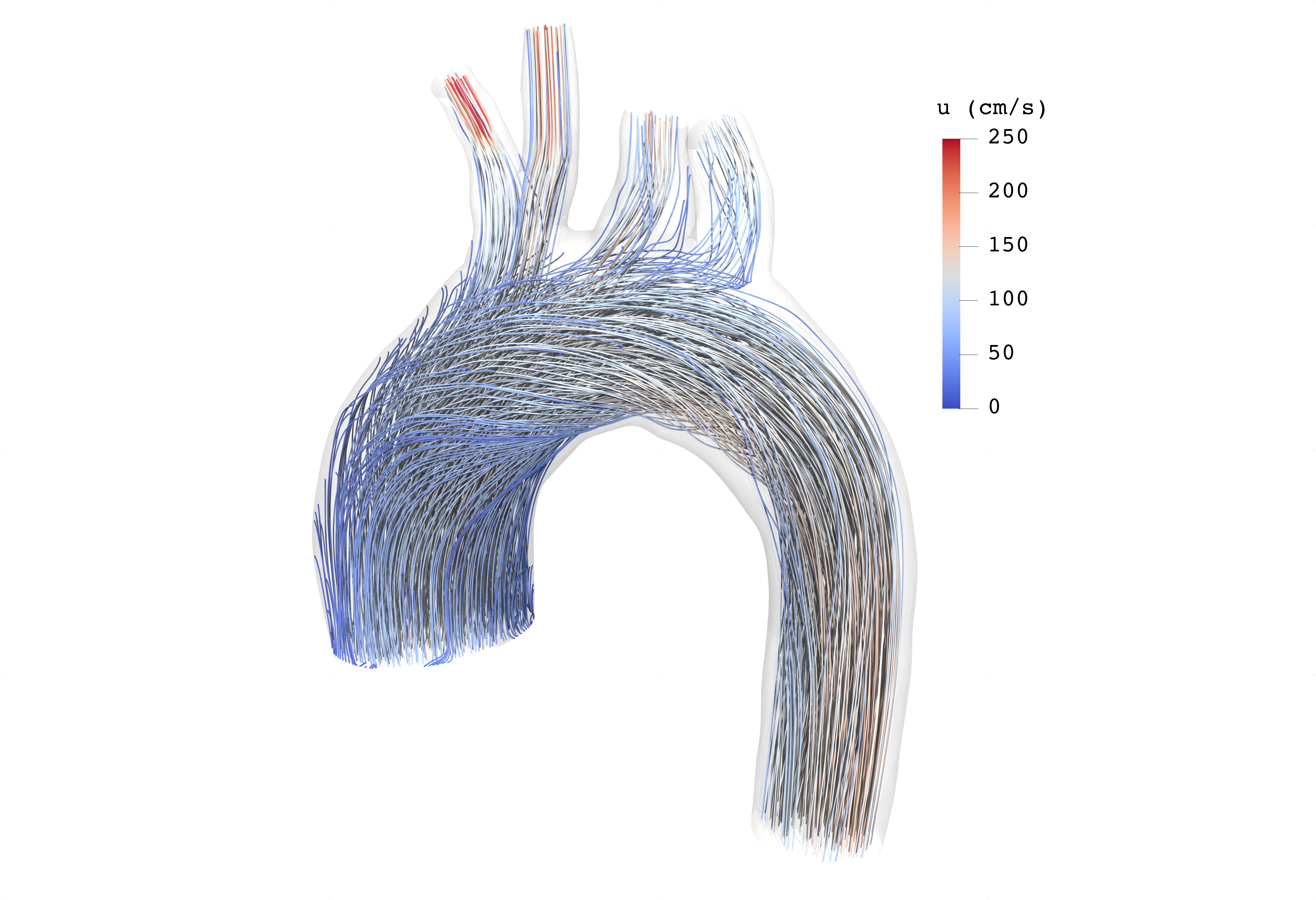} }
  \subfloat[\footnotesize{FEM with estimated \\ params. ($t=0.12$ s)}]{
  \includegraphics[trim=1000 20 900 20, clip, width=0.29\textwidth]{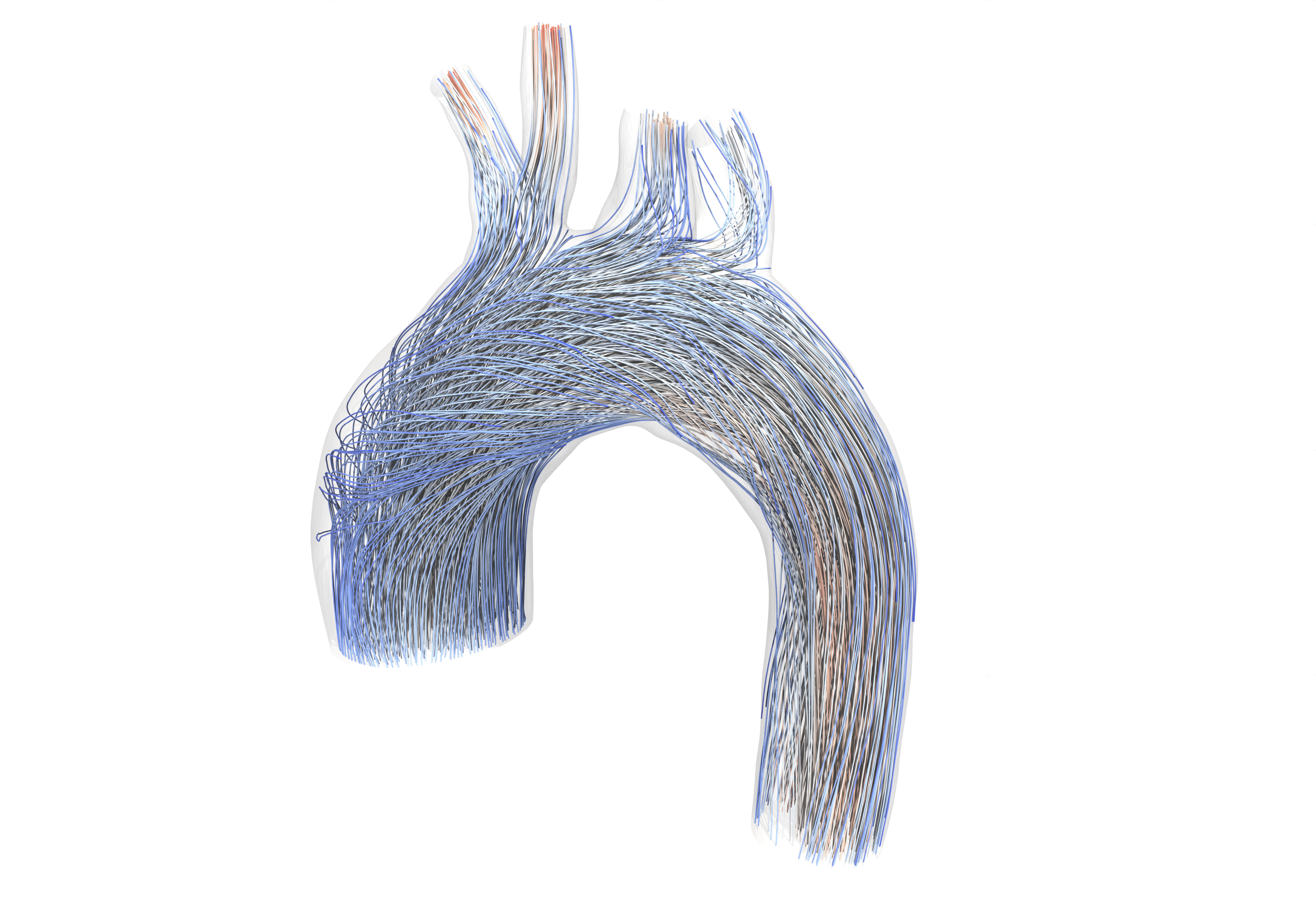} } \\
  \subfloat[\footnotesize{Reference ($t=0.45$ s)}]{
  \includegraphics[trim=900 20 900 20, clip, width=0.3\textwidth]{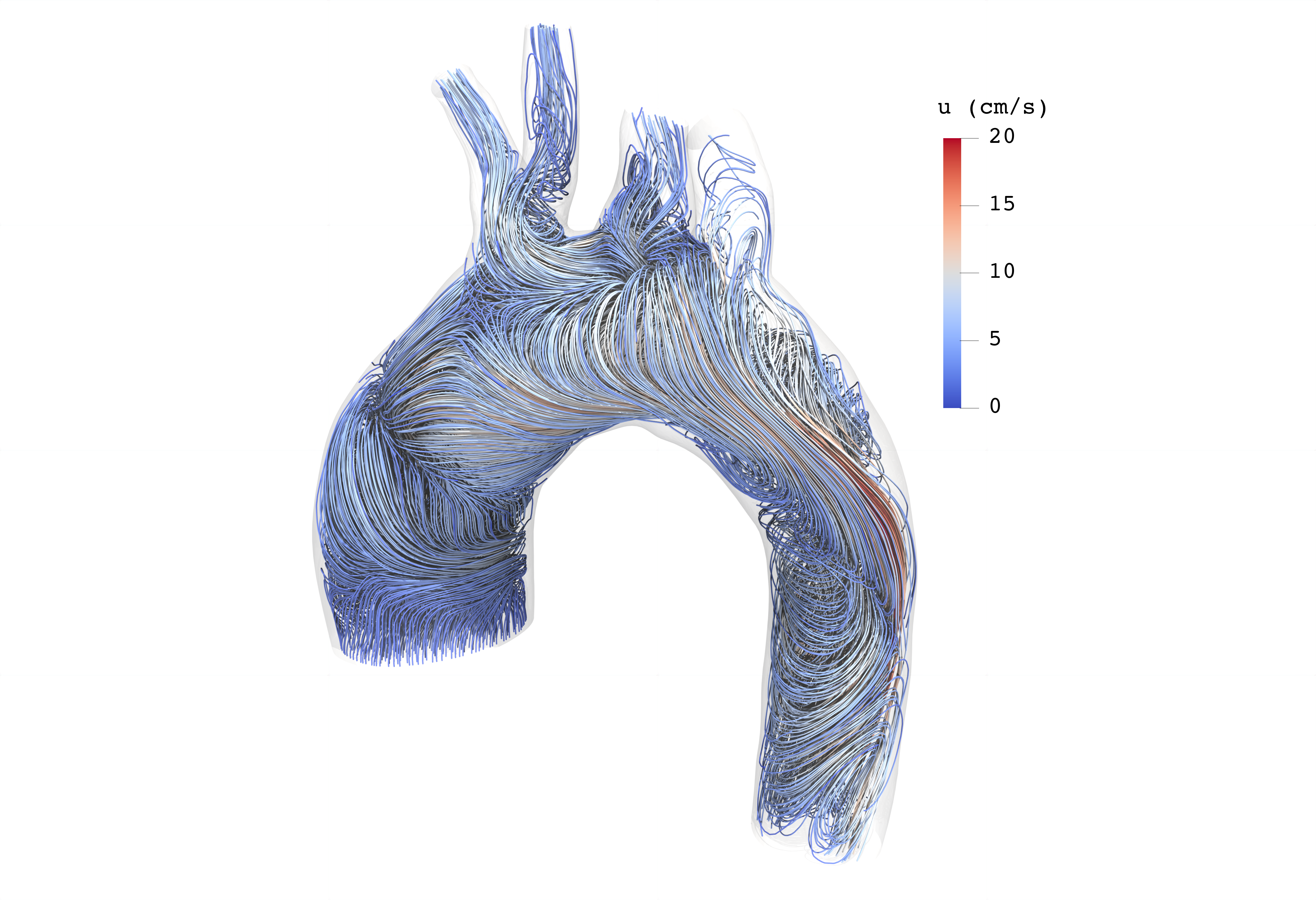} } 
\subfloat[\footnotesize{PINNs ($t=0.45$ s)}]{
  \includegraphics[trim=1000 20 900 20, clip, width=0.29\textwidth]{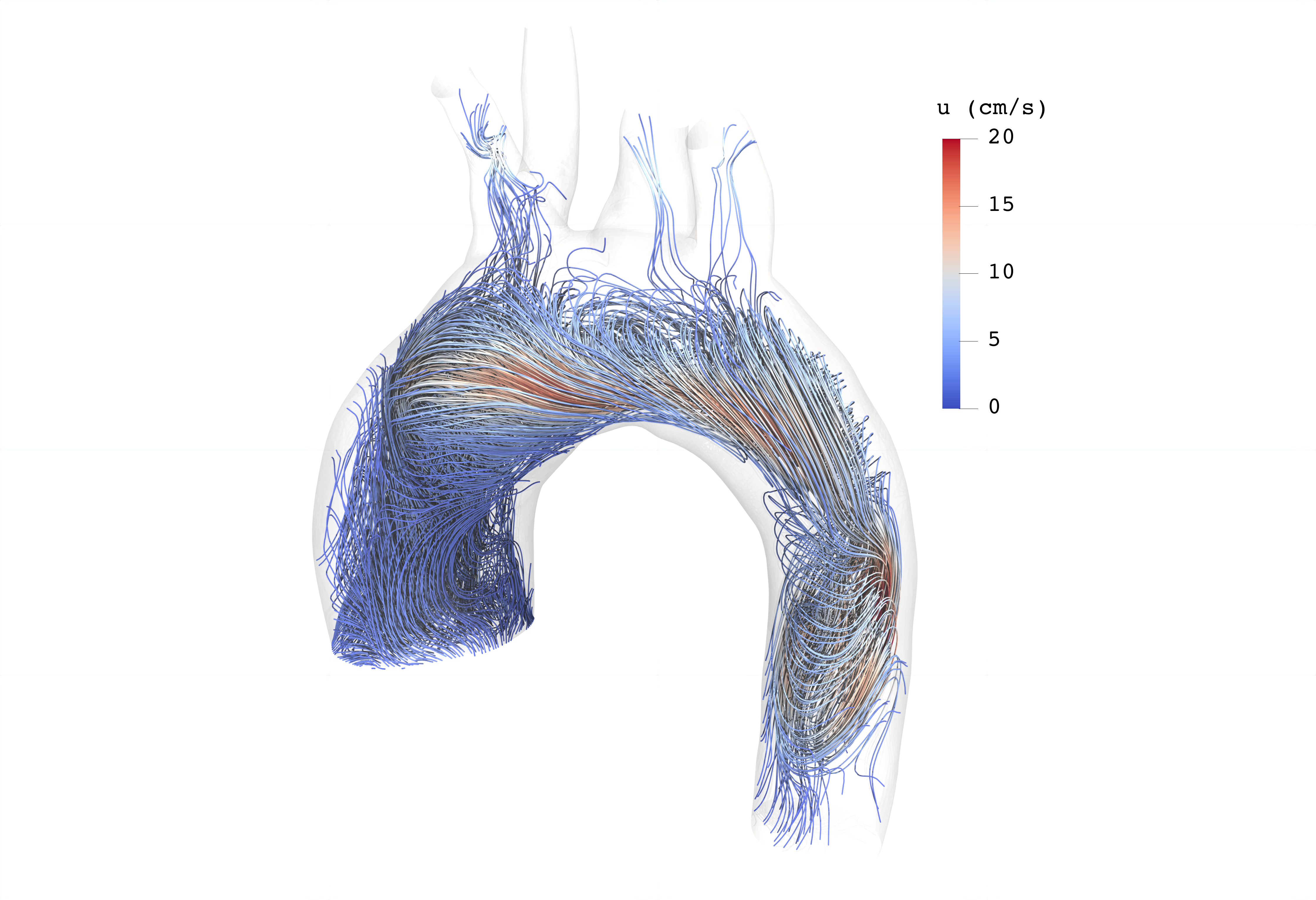} }
  \subfloat[\footnotesize{FEM with estimated \\ params. ($t=0.45$ s)}]{
  \includegraphics[trim=1000 20 900 20, clip, width=0.29\textwidth]{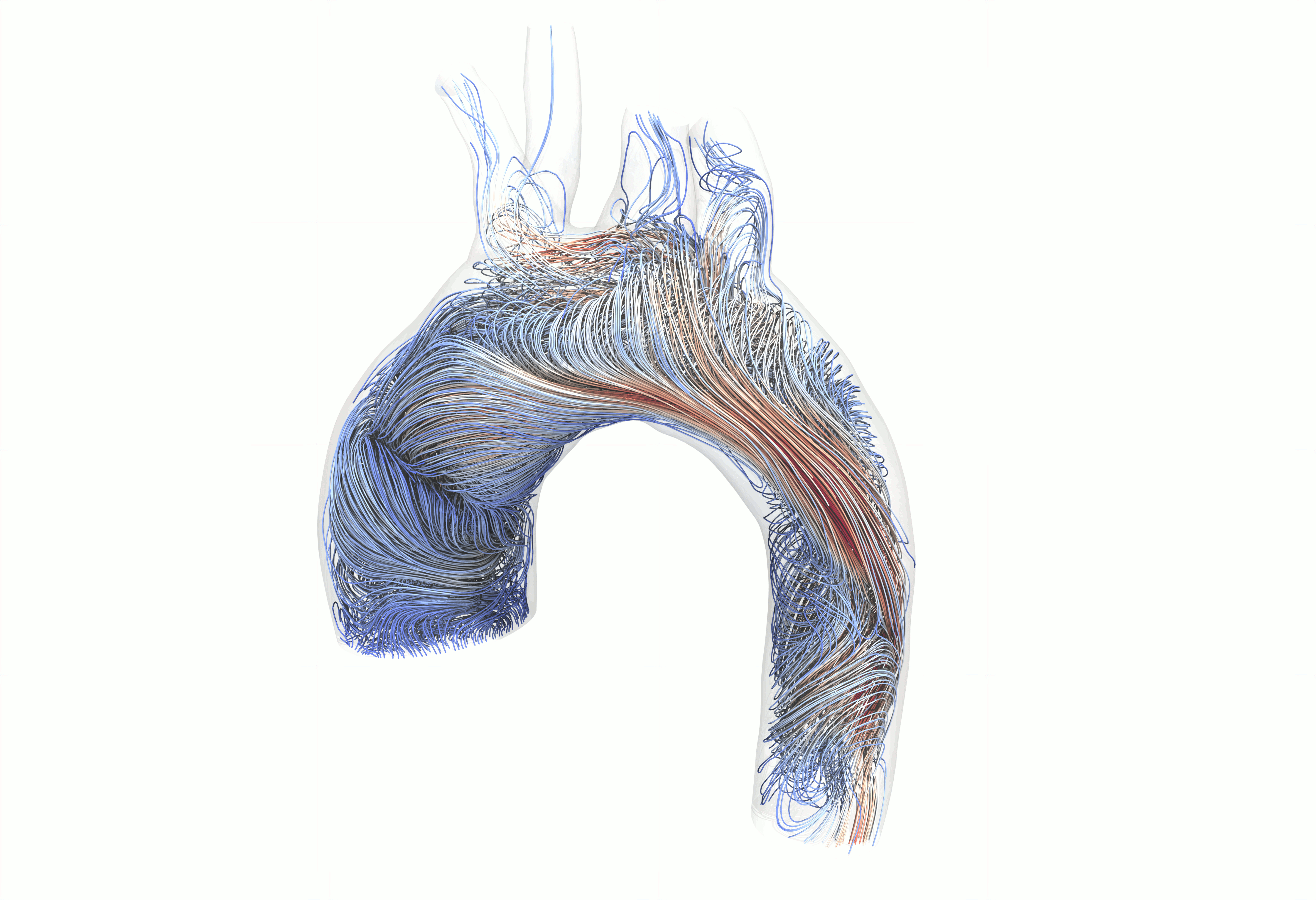} }
  \caption{Velocity streamlines of the transient problem for the reference solution (a and d), the mean estimated velocity estimated by the PINNs (b and e), and the velocity obtained after a FEM simulation using the found parameters (c and f) at two time instants: $t=0.12 \ s$, which correspond to peak systole, and at $t=0.45 \ s$, during mid diastole.}
  \label{fig:streamlines_transient}
\end{figure}

Finally, streamlines computed from the nodes of $\Gamma_{in}$ were calculated at two instants of the cardiac cycle and shown in Figure \ref{fig:streamlines_transient}. This was done for the reference velocity, the PINN's estimation, and the FEM post-process simulated velocity. Generally, the neural network can capture the main flow features at systole but not so well during diastole, where the velocities are lower and the measurements' velocity-to-noise ratio is worse. When comparing the reference velocity against the FEM post-process solution, no significant gain in quality can be seen while in systole. However, during diastole, the FEM solution captured the main flow features that the PINNs' estimation failed to represent, especially the vortices at the ascending aorta and the near-wall velocity in the entire domain.

\section{Discussion}
\label{sec:discussion}

In this we work, we develop a novel method to estimate hemodynamical parameters from sparse measurements using PINNs. We note that the original problem was divided into two different flow regimes: stationary and transient. We made this split to study the quality of the results when the problem grows in complexity.

The stationary problem could be seen as a inexpensive version of the transient problem, and in clinical hemodynamics, is usually studied when the transient problem's computational time exceeds the allowed limit imposed by the clinical application itself \cite{vinoth2019steady, ene2014influence}. In terms of the quality of the estimations, it could be seen from the results in Table \ref{tab:steady_results} that although all parameters were overestimated, the obtained values present a mean error of less than $2\%$ when comparing with their reference. On the other hand, when looking into the reconstructed velocity streamlines in Figure \ref{fig:streamlines_steady}, both reconstructed and estimated fields are very similar, with some mild differences at the vessel wall. One of the reasons for that difference is because FEM methods exactly imposes the boundary condition on the walls, being the velocity there set exactly to zero, as it happened with the reference solution. However, in the case of the PINNs method presented here, the non-slip condition enters the optimization as a loss term and, consequently, is not completely satisfied. This produces some of the velocity lines to go out of the vessel through the walls. This effect is also present in the transient flow regime.

It is particularly interesting that while the velocity tends to be misrepresented near the vessel wall, the outflow values are not. We hypothesized that since the coupling between the Navier-Stokes and Windkessel model is in terms of pressures and flows \cite{windkesselartery}, the extra flow condition in the total loss in Equation \eqref{ec:loss_wk} acts as a reinforcement for the training. Consequently, the outlet velocity tends to rise in the lumen, compensating for the low velocity near the wall. This effect is increased during the transient diastole, as some of the upper aortic outlets do not perceive any velocity, as shown in Figure \ref{fig:transient_flows}(b,e). One strategy to mitigate this effect could be introducing a hard-constrain for the non-slip condition into the optimization, as done for a 2D Navier-Stokes system in Ref \cite{gao2021super}. Another strategy could be relaxing the wall definition by introducing a \emph{slip/transpiration} condition as it was done in Nolte et al. \cite{nolte2019reducing} when studying the impact of segmentation's errors in MRI blood flow measurements.

Another aspect to discuss more in detail is the use of the potential vector representation for the flow velocity in Equation \eqref{ec:phi_def}. This strategy could be seen as a hard constraint, applied only to the transient problem, to exactly satisfy the mass-conservation term of the Navier-Stokes model, as it was also done in Ref \cite{mohan2020embedding} for turbulent flows. We found that this change of variables improved the overall estimation of the method while increasing the computational time by a factor of 4. Also, it is known that the change of variables in Equation \eqref{ec:phi_def} present a \emph{gauge freedom}, meaning that any exact gradient of the form $\nabla \psi$ added to $\Phi$, will end in the same velocity profile. In this work, we did not explore how different gauges could impact the inverse problem performance. Future work may include studying those and how they can help enhance the properties of the optimization method \cite{marner2019potential}. 

The present methodology is not free of limitations and has to be seen as a first step in using neural networks for solving clinical real-world inverse problems. One of the limitations of this work is the lack of experiments with real clinical data. In this line, future work includes the use and assimilation of real patient MRI blood flow images, which we expect will need better strategies in how to deal with such data, as previous work in \cite{garay2022parameter} and in \cite{fathi2020super} have done to mitigate the presence of artifacts and poor quality of the images.

Finally, since the flexibility in defining a forward and inverse problem within the PINNs workflow, this method could be easily extended to other clinical-relevant tasks such as the denoising and undersampling of MRI blood flow measurements as it was done in \cite{mura2016enhancing,garay2020new,partin2023analysis} with high dependence on a numerical solver and model assumptions. In this line, the PINNs workflow could be seen as a good candidate with less restriction in using contaminated low-resolution data.

\section{Conclusions}
\label{sec:conclusions}

In this work, we present how a physics-informed neural network (PINNs) can be used for estimating hemodynamical information from MRI-like images in a synthetic scenario, where a reduced-order model was coupled with the 3D Navier-Stokes equations to generate the blood flow images. The PINNs were able to solve the coupled system in two regimes: steady and transient flow, using simulated 2D MRI images, a mean pressure curve, the non-slip condition at the wall, and the model equations. Furthermore, at the end of the training, the algorithm was able to estimate the best matching physical parameters, which in this case, represent the elastic response of the adjacent vasculature of the study domain. The overall result shows good accuracy when compared with the reference values in the steady regime and moderately worsened results when dealing with the transient state, where the number of parameters is increased by three-fold, and the data dimensionality is also increased.

To summarize, the proposed framework provides an initial step to truly personalize hemodynamic models from clinical data using a neural network workflow. We envision patient-specific models enabling better diagnostics and treatments for various cardiovascular diseases.

\section*{Acknowledgements}
This work was funded by ANID Chile: Millennium Science Initiative Program ICN17\_002 (IMFD), ICN2021\_004 (iHealth) and NCN19\_161 (ACIP), Fondecyt grant 11201250, FONDECYT-Iniciaci\'on 11220816, Basal Funds for Center of Excellence FB210005 (CMM) and Fondecyt Postdoc grant 3230549.



\bibliographystyle{elsarticle-num}

\bibliography{TheOP.bbl}

\end{document}